\documentclass[11pt]{article}
\usepackage{epsfig}
\usepackage{graphicx}
\usepackage{amssymb}
\usepackage[mathscr]{euscript}
\usepackage{amsmath}
\usepackage{oldgerm}
\usepackage{picinpar}
\usepackage{theorem}

\def\integer{{\mathbb{Z}}}

\def\real{{\mathbb{R}}}

\def\proba{{\rm I\kern -.18em P}}

\newcommand{\I}{{i}}
\newcommand{\D}{{\rm{d}} }

\newcommand{\dc}{{\mathcal D}}

\newcommand{\nc}{{\mathcal N}}
\newcommand{\lc}{{\mathcal L}}
\newcommand{\ooc}{{\mathcal O}}
\newcommand{\pc}{{\mathcal P}}
\newcommand{\tc}{{\mathcal T}}
\newcommand{\ssc}{{\mathcal S}}
\newcommand{\qv}{\mbox{\boldmath{$q$}} }
\newcommand{\pv}{\mbox{\boldmath{$p$}} }
\newcommand{\xv}{\mbox{\boldmath{$x$}} }
\newcommand{\yv}{\mbox{\boldmath{$y$}} }
\newcommand{\zv}{\mbox{\boldmath{$z$}} }
\newcommand{\Tv}{\mbox{\boldmath{$T$}} }
\newcommand{\Nv}{\mbox{\boldmath{$N$}} }

\newcommand{\heff}{{\hbar_{\rm{eff}}} }

\begin{document}



\title{Spectral form factor of hyperbolic systems: leading off-diagonal approximation}
\author{D.~Spehner\\
{\it Universit\"at Duisburg-Essen, Fachbereich Physik,
D-45117 Essen, Germany} \\
E-mail: spehner@theo-phys.uni-essen.de
}
\date{}
\maketitle

\begin{abstract}
The spectral fluctuations of a quantum Hamiltonian system with time-reversal symmetry
 are studied
in the semiclassical limit by using periodic-orbit theory.
It is found that, if long periodic orbits
are hyperbolic and  uniformly distributed
in phase space, the spectral form factor 
$K(\tau)$ agrees with the GOE prediction of 
random-matrix theory up to second order included in the time $\tau$ measured in units of the 
Heisenberg time (leading off-diagonal approximation).
Our approach is based on the mechanism of periodic-orbit correlations 
 discovered recently by  Sieber and 
Richter~\cite{Sieber01}. By reformulating  the theory of these authors in phase space,
their result on the free motion on a Riemann surface with constant negative curvature is extended to 
general Hamiltonian hyperbolic systems with two degrees of freedom.
\end{abstract}


\section{Introduction}

One of the fundamental characteristics of quantum systems with classical chaotic dynamics
is the universality of their spectral fluctuations. This universality and the 
agreement with the predictions of random-matrix theory (RMT)
was first conjectured 
 by Bohigas, Giannoni and Schmit (BGS)~\cite{BGS84}. It has been later supported
by numerical investigations on a great variety of systems~\cite{Haake}.
However, the necessary and sufficient conditions on the underlying 
classical dynamics leading to such a universality in quantum spectral statistics are not known, 
and  the origin of the success of RMT in clean chaotic systems is still subject to debate.

In the semiclassical limit, where the BGS conjecture is expected to be valid,
the Gutzwiller trace formula~\cite{Gutzwiller} expresses the density of states  
$\rho(E) = \sum_{n} \delta (E - E_n)$ of the quantum system
as a sum of a smooth part $\overline{\rho}(E)$ and an oscillating part. The latter  
is given by a sum ${\rho}_{\text{osc}} (E) 
 = (\pi \hbar)^{-1} \sum_\gamma A_\gamma \,\cos ( S_\gamma/\hbar - \pi \mu_\gamma/2 )$
 over all classical periodic orbits $\gamma$ of energy $E$
($S_\gamma$ and $\mu_\gamma$ are the action and the Maslov index of $\gamma$, 
and $A_\gamma$ is an associated amplitude). 
The energy correlation function, 
\begin{equation} \label{eq-R(E)}
R (\epsilon)  =  \frac{1}{\overline{\rho}(E)^{2}} \left\langle \rho 
 \left( E + \frac{\epsilon}{2} \right) 
 \rho \left( E - \frac{\epsilon}{2} \right) \right\rangle_E -1\;,
\end{equation}
and its Fourier transform $K(\tau)$, the so-called form factor,
are given  by  sums over {\it pairs} $(\gamma,\gamma')$ of 
periodic orbits. Here  $\tau$ is the time 
measured in units of the Heisenberg time
$T_H =  2 \pi \hbar \,\overline{\rho} (E)$ 
($T_H = \ooc(\hbar^{1-f})$ for systems with $f$ degrees of freedom).
The brackets denote an (e.g. Gaussian) energy average over an energy 
width $W$ much larger than  the mean level spacing
$\Delta E = \overline{\rho} (E)^{-1}$, but  classically  small, $W \ll E$,
so that $\langle \rho \rangle_E \simeq \overline{\rho}(E)$. 
By neglecting the `off-diagonal' terms, i.e., 
the contributions of pairs   
of distinct orbits mo\-dulo symmetries, 
Berry~\cite{Berry85} showed that  the  spectral fluctuations of classically chaotic 
systems  
agree in the limit $\hbar \rightarrow 0$ with the RMT predictions to first order in $\tau$
($\tau \ll 1$).
Two different approaches have been proposed to support the BGS conjecture 
to all orders in $\tau$ in the semiclassical limit.
The first one is 
based on a mapping between the parameter level dynamics and   the dynamics
of a  gas of fictitious particles~\cite{Haake,Pechukas83}.  
The second one uses  field-theoretic and supersymmetric
methods and applies to
systems with  exponential decays of  classical correlation functions~\cite{Agam95}. 

The link between spectral correlations and correlations among periodic orbits
 was first put forward  in~\cite{Argaman93}. It was argued in this reference  
that the BGS conjecture implies some universality at the level of 
classical action correlations.
Recently, Sieber and Richter~\cite{Sieber01} identified a general mechanism leading to
correlations among periodic orbits in chaotic systems 
with two degrees of freedom having a time-reversal  invariant dynamics. 
This has opened the route
towards an understanding of the universality of spectral fluctuations based on periodic-orbit theory
only. The crucial fact is 
that an orbit $\gamma$ having  a self-intersection  
in configuration space with nearly antiparallel velocities
is correlated with another orbit $\tilde{\gamma}$, having an avoided intersection instead of a 
self-intersection, which
has almost the same action  and amplitude.
In two special systems, the free motion on a Riemann surface
 with constant negative curvature (Hadamard-Gutzwiller model)~\cite{Sieber01}
and quantum graphs~\cite{Schanz02},
the pairs $(\gamma, \tilde{\gamma})$ have been  found to give a contribution 
$K_2(\tau) = -2 \tau^2$ to the semiclassical form factor. This result is in agreement with 
the Gaussian orthogonal ensemble (GOE) prediction of RMT,
\begin{eqnarray} \label{eq-K_GOE}
\nonumber
K_{\text{GOE}}(\tau)  
& =  &  2 \tau - \tau \ln ( 1 + 2 \tau ) \;\;\;, \;\;\; 0 < \tau < 1
\\  
& = &  2 \tau - 2 \tau^2 + \ooc (\tau^3)  \;.
\end{eqnarray}
The first term $K_1(\tau)=2 \tau$ is obtained by using Berry's diagonal approximation.

The purpose of this work is to extend Sieber and Richter's result
to general hyperbolic and ergodic two-dimensional Hamiltonian systems. 
Unlike in~\cite{Sieber01}, our approach does not rely on the concepts of self-intersections 
and avoided intersections with nearly antiparallel velocities,   
but rather focus on what corresponds to such events in phase space, namely 
{\it the existence of two stretches of the orbit} (for both $\gamma$ and $\tilde{\gamma}$) 
{\it which are almost time reverse of one another}.
It will be argued that  working in phase space 
has a number of  advantages and may allow 
for easier generalisations to periodically driven systems
and to systems with $f>2$ degrees of freedom. 
A similar approach is presented in~\cite{Marko}; an alternative approach, based 
on a projection onto the configuration space as in~\cite{Sieber01}, is presented in~\cite{Mueller03}.

In section~\ref{sec-hyp}, we state the main hypothesis on the classical dynamics 
used throughout this paper. After having briefly recalled the main ingredients of 
the theory of Sieber and Richter in section~\ref{eq-RS_theory}, 
a characterisation
of the orbit pairs $(\gamma,\tilde{\gamma})$  in the Poincar\'e surface of section
is given in section \ref{sec-phase_sp}.
The unstable and stable coordinates associated with a pair $(\gamma,\tilde{\gamma})$
are introduced in the following section. The leading off-diagonal
correction $K_2(\tau)=-2 \tau^2$ to the semiclassical form factor is derived in 
section~\ref{sec-form_fact}. Our conclusions are drawn in the last section. Some 
technical details are presented in two appendixes.

\section{Hyperbolic Hamiltonian systems} \label{sec-hyp}

\begin{center}
\begin{figure}
\centering
\includegraphics[width=8cm]{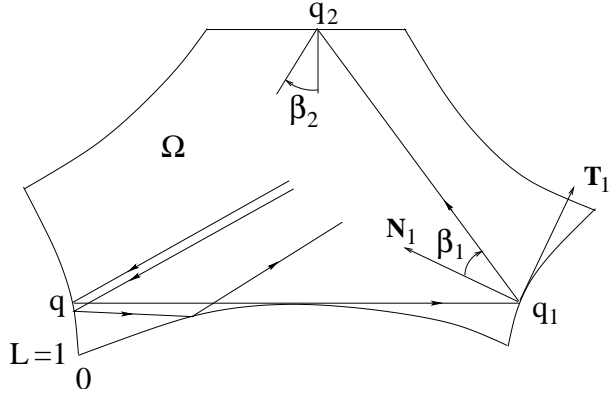}
\caption{\it Billiard $\Omega$: $q_n$ and $p_n = \sin\beta_n$ are the arc length along $\partial \Omega$
and the momentum tangential to $\partial \Omega$ in dimensionless 
units.}
\label{fig-0}
\end{figure}
\end{center}

We consider a particle moving in a Euclidean plane ($f=$ two degrees of freedom), 
with  Hamiltonian $H(\qv,\pv)=H(\qv,-\pv)$ 
invariant under time-reversal symmetry.  
We assume  the existence of a compact two-dimensional Poincar\'e surface of section 
$\Sigma$ in the (four-dimensional) phase space $\Gamma$, contained in an energy shell 
$H(\qv,\pv) = E$
and invariant under time reversal (TR)~\cite{Gutzwiller,Gaspard}.
Every classical orbit of energy $E$ intersects $\Sigma$ transversally. 
The classical dynamics can then be described by an area-preserving 
map $\phi$ on $\Sigma$,  together with
a first-return time map $x \in \Sigma \mapsto t_x \in [0,\infty]$ (see \cite{Gaspard}). 
In what follows, letters in normal and bold fonts are assigned to the canonical coordinates 
$x=(q,p)$ in $\Sigma$ and to points $\xv=(\qv,\pv)$ in $\Gamma$, respectively.
It is convenient to use dimensionless $q$ and $p$ by measuring them
in units of some reference length  $L$ and  
momentum $P$.  
The $n$-fold iterates of $x$ by the map are denoted by $x_n = \phi^n\,x$, $n \in \integer$. They
are the coordinates of the intersection points $\xv_n$ of a 
phase-space trajectory with $\Sigma$, according to a given direction of traversal.
The Euclidean distance between two points of coordinates $x$ and $y$ in $\Sigma$ is denoted by 
$| y - x|$. 
If the system is a billiard ($H(\qv,\pv) = \pv^2/2M$ if $\qv$ is inside a compact domain 
$\Omega \subset \real^2$ and $+\infty$ otherwise), $\Sigma$ is 
the set of points $(\qv,\pv) \in \Gamma$ such that $\qv$ is on the boundary $\partial \Omega$
of the billiard, $\pv$ is the momentum after the reflection on $\partial \Omega$,  
and $|\pv | = \sqrt{2 M E}$. Then
$q$ is the arc length along $\partial \Omega$ 
in units of the perimeter $L$, 
$p$ is the momentum tangential to $\partial \Omega$
in units of $\sqrt{2 M E}$, and $t_{x}$ is 
the length of the segment of straight line linking two consecutive reflection  points,
multiplied by the inverse velocity $\sqrt{M/2E}$ (see Fig.~\ref{fig-0}). 
Due to the Hamiltonian nature of the dynamics,
the linearised $n$-fold iterated map $M_x^{(n)} = D_x ( \phi^n)$ 
is symplectic. This means that it conserves the symplectic product
\begin{equation} \label{eq-symplec_prod}
\Delta x \wedge \Delta x'
  = 
   \Delta p \, \Delta q ' - \Delta q \, \Delta p '
\end{equation}
for any two infinitesimal displacements $\Delta x= (\Delta q, \Delta p)$ and 
$\Delta x' = (\Delta q' , \Delta p')$ in the tangent space $\tc_x \Sigma$.

The time reversal (TR) acts
in the phase space $\Gamma$ by changing the sign of the momentum, 
$T_\Gamma: (\qv,\pv) \mapsto (\qv,-\pv)$. 
Its action on $x$ is given by an area-preserving 
self-inverse map $T$.
When acting on an infinitesimal displacement $\Delta x$ in $\tc_x \Sigma$,
the same symbol $T$ refers to the linearised version of $T$ (we avoid 
the cumbersome notation $D_x T$, the meaning of $T$ being clear from the context). 
In most cases, the exact map $T$ is   already linear and given by
$T: (q,p) \mapsto  (q,-p)$.
The TR symmetry of the Hamiltonian implies 
$\phi\, T = T \,\phi^{-1}$, i.e., $(Tx)_n = T x_{-n}$.

Some spatially symmetric
systems in an external  magnetic field have 
non-conventional TR symmetries, obtained
by composing $T_\Gamma$ with a canonical 
transformation associated with the spatial symmetry~\cite{Haake}. 
The Sieber-Richter pairs $(\gamma,\tilde{\gamma})$ of
 correlated orbits also exist in such systems, 
although they look different in configuration space~\cite{Braun02}.  
By performing the canonical transformation to
redefine new coordinates $(\qv,\pv)$  at the beginning, 
the TR becomes the conventional
one. Therefore  
the analysis below also applies to systems with non-conventional TR symmetries.

The normalised $\phi$-invariant measure is the Liouville measure
$\D  \mu(q,p) = \D q  \, \D p /| \Sigma|$, where
$|\Sigma |$ is the (dimensionless) area of $\Sigma$.
Our main assumptions on the dynamical system $(\phi,\Sigma,\mu)$ are
\begin{itemize}
\item[(i)] $\mu$ is ergodic; 
\item[(ii)] all  
Lyapunov exponents  are different from zero
on a set of points $x$ of measure one (complete hyperbolicity); 
\item[(iii)]
long periodic orbits are `uniformly  distributed in $\Sigma$'.
\end{itemize}
Note that (i) and (ii) imply that the Lyapunov exponents $\pm \lambda_x$
are constant $\mu$-almost everywhere
and equal to $\pm \langle \lambda \rangle$, with $\langle \lambda \rangle  > 0$
(the periodic points are notable exceptions of measure zero where this is wrong!).
Examples of billiards satisfying (i-ii) are  
semi-dispersing billiards (if trajectories reflecting solely on the neutral part of 
$\partial \Omega$ form a set of measure zero), 
the  stadium and other Bunimovich billiards, the cardioid billiard, and the periodic Lorentz 
gas (see e.g.~\cite{Chernov96} and references therein).  
Assumption (iii), associated with ergodicity (i), means that 
an (appropriately weighted) average over periodic orbits with periods inside a given time window $[T,T+\delta T]$
can be replaced in the large-$T$ limit by a 
phase-space average~\cite{Bowen72,Knieper98}. Note that this statement, which is the precise
content of (iii), does not concern individual periodic orbits but rather
averages over many periodic orbits with large periods. 
We think that the statement can hold true even if some periodic orbits with arbitrary large periods
are stable, if there are exponentially less such orbits than unstable orbits. 

As is typically the case 
in billiards, the Poincar\'e map $\phi$ or its derivatives may be 
singular on a closed set $\ssc \subset \Sigma $ of measure zero. 
For instance, if the  boundary $\partial \Omega$ is concave outward, $\phi$ is discontinuous at
a point $x_S = (q_S,p_S)$ such that the 
trajectory between $\qv_S$ and the next reflection point  
is tangent to $\partial \Omega$  at this point (see Fig.~\ref{fig-0}).   
Let us denote by $d(x,\ssc)$ the Euclidean 
distance from $x$ to $\ssc$. We assume that
\begin{itemize}

\item[(iv)] $\ssc$ is `not too big':
$\mu ( B_{\delta,S} ) \leq C_1 \,\delta^{\sigma_1}$ for any $\delta>0$, with 
$B_{\delta,S} = \{ x \in \Sigma; | x - x_S | \leq \delta, x_S \in \ssc \}$ and $\sigma_1 >0$;

\item[(v)] 
the divergence of the derivatives of $\phi$ on $\ssc$ is at most algebraic,
$| \partial^r \phi/\partial x^{\alpha_1} \ldots \partial x^{\alpha_r} | \leq 
C_2\,d( x, \ssc)^{-\sigma_2(r-1)}$, with  $\sigma_2 >0$. 

\end{itemize}
Here $C_1 >0$ and $C_2 \geq 1$ are constants of order one
and the
 indices $\alpha,\beta=1,2$  refer to the $q$- and $p$-coordinates in $\Sigma$
($x^1 = q$, $x^2 = p$).
(iv-v) are standard mathematical assumptions on billiard maps~\cite{Katok}.


\section{The theory of Sieber and Richter} \label{eq-RS_theory}

The starting point of Sieber and Richter is the semi-classical expression of
the form factor, 
\begin{equation} \label{eq-K_semicl}
K_{\text{semicl}} \left( \tau = \frac{T}{T_H} \right) 
 = \frac{1}{T_H} \, 
  \frac{1}{\delta T} 
   \left\langle \;
\sum_{T \leq (T_{\gamma}+T_{\gamma'})/2 \leq T+\delta T} 
A_{\gamma} A_{\gamma'} \,
    e^{ \frac{\I}{\hbar}(S_{\gamma} -S_{\gamma'}) 
     - \I \frac{\pi}{2} (\mu_{\gamma} - \mu_{\gamma'} )} 
\right\rangle_E\;.
\end{equation}
The sum runs over all pairs of periodic orbits $(\gamma,\gamma')$ such that  the half-sum of 
their periods
$(T_{\gamma}+T_{\gamma'})/2$ is in the time window 
$[T, T + \delta T]$ of width $\delta T \ll T_H$.
For isolated periodic orbits, 
$A_\gamma = T_\gamma \,r_\gamma^{-1} | \det (M_\gamma^{(F)} -1 ) |^{-1/2}$, where $r_\gamma$ is
the repetition (number of traversals) of $\gamma$
and  $M_\gamma^{(F)}$ is the stability matrix of $\gamma$ for displacements perpendicular 
to the motion~\cite{Gutzwiller}. 
In order to work with a self-averaging form factor~\cite{Prange97}, a time averaging over the window 
$[T, T + \delta T]$ (with, e.g., $\delta T = h/W$)
has been performed in (\ref{eq-K_semicl}).
Equivalently, $K (\tau)$ can be defined as the truncated Fourier transform
\begin{equation}
K (\tau)  = 
\overline{\rho}(E) \int_{-\infty}^\infty  \D \epsilon \,
 R(\epsilon)\,
\frac{\sin  \left( \epsilon \,\delta T/2 \hbar \right) }{\epsilon \,\delta T/2 \hbar}
 \, e^{-\frac{\I}{\hbar} \epsilon T }
\end{equation}
 of the
energy correlation function (\ref{eq-R(E)}). Formula
(\ref{eq-K_semicl}) gives the correct form factor for small enough times $\tau$ only. It
relates the {\it quantum energy correlations} 
to the {\it classical action  correlations}~\cite{Argaman93}. 
Indeed, only  orbits with  correlated actions, differing by an amount of order $\hbar$,
can interfere constructively in (\ref{eq-K_semicl}).

\begin{center}
\begin{figure}
\centering
\includegraphics[width=12cm]{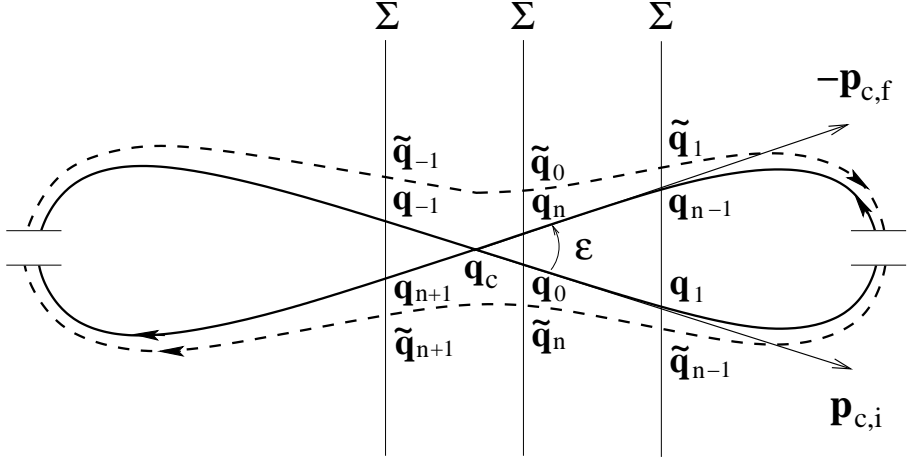}
\caption{\it Pairs of correlated periodic orbits $\gamma$ (solid line)
and $\tilde{\gamma}$ (dashed line) in configuration space for systems with conventional TR symmetry. 
The intersections of $\qv$-space with 
$\Sigma$ are schematically represented by parallel vertical lines.}
\label{fig-1}
\end{figure}
\end{center}

For fixed $\tau = T/T_H >0$, the sum (\ref{eq-K_semicl}) deals with orbits 
with very long periods as $\hbar \rightarrow 0$ (recall that $T_H = \ooc (\hbar^{-1})$). 
Such orbits have many 
self-intersections in $\qv$-space, some of them characterised by small  angles $\varepsilon$
at the crossing point.  
As shown in~\cite{Sieber01}, the two loops at both sides of the crossing point can be 
slightly deformed in  such a way that they form a neighbouring closed orbit in $\qv$-space, having 
 an avoided crossing instead of a crossing (see Fig.~\ref{fig-1}).
The two partner orbits $\gamma$ and $\tilde{\gamma}$ are almost time reverse of 
each other on one loop (right loop) and almost coincide on the other 
(left loop). 
Such a construction, which was supported in~\cite{Sieber01} 
by using the linearised dynamics, is in general possible   in systems with TR symmetry 
and for small enough $\varepsilon$  only.
Due to the hyperbolicity of $\gamma$, the two orbits come
exponentially close to each other in $\qv$-space as 
one moves away from the crossing point $\qv_c$. 
This means that the 
phase-space displacement perpendicular to the motion associated with $\gamma$ and 
 $\tilde{\gamma}$ is  almost (but not exactly) on the unstable  
manifold of $\gamma$ at $\xv_{i,c}=(\qv_{c},\pv_{c,i})$, whereas 
the displacement  associated with $\gamma$  
and the TR of $\tilde{\gamma}$  is almost on the stable manifold  of $\gamma$ at 
$\xv_{i,c}$~\cite{Braun02}.
If a symbolic dynamics is available, the  symbol sequence of $\tilde{\gamma}$ 
can be constructed from the symbol sequence of $\gamma$ in a simple way~\cite{Heusler02};
in the Markovian case, the TR symmetry implies that the partner sequence must not be pruned. 
Since the two orbits $\gamma$ and $\tilde{\gamma}$ have almost the same period 
and almost the same Lyapunov exponents, the amplitudes $A_\gamma$ and $A_{\tilde{\gamma}}$
are almost equal. Furthermore, it can be
shown by using a winding number argument 
that $\mu_{\tilde{\gamma}} = \mu_\gamma$~\cite{Marko,Mueller03}.
In the Hadamard-Gutzwiller model, the difference $\delta S$ of the actions of 
$\tilde{\gamma}$ and $\gamma$ is given by $\delta S \simeq E \,\varepsilon^2/ \lambda^{(F)}$  
in the small $\varepsilon$ limit, where $\lambda^{(F)}$ is the positive Lyapunov exponent 
of the Hamiltonian flow~\cite{Sieber01}.  
The main hypothesis of~\cite{Sieber01} is that, if the system has
no other symmetries than TR, only the pairs $(\gamma,\tilde{\gamma})$ contribute 
 to the leading off-diagonal correction $K_2(\tau)$ to the semiclassical form factor 
(\ref{eq-K_semicl}) in the limit $\hbar \rightarrow 0$,
\begin{equation} \label{eq-K2}
K_2(\tau)  
 = 
\frac{1}{T_H} \, 
  \frac{1}{\delta T} 
   \left\langle \;
\sum_{T \leq  T_\gamma \leq T+\delta T} 
A_\gamma^2 \,
 \sum_{\text{$\tilde{\gamma}$ partner of $\gamma$}} e^{\frac{\I}{\hbar} \, \delta S} 
\right\rangle_E\;.
\end{equation}
The main task is to evaluate the right-hand side. This was performed up to now for the 
Hadamard-Gutzwiller model~\cite{Sieber01} and for quantum graphs~\cite{Schanz02}. 
The main difficulties 
arising in extending the theory of  Sieber and Richter to other systems satisfying the 
hypothesis in the previous section are

\begin{itemize}

\item the orbit $\gamma$ may have a family of correlated self-intersections, 
corresponding to one and the same
partner orbit $\tilde{\gamma}$; this happens for instance in focusing 
billiards~\cite{Mueller03,Mueller}; care must be taken to avoid
overcounting the pairs $(\gamma,\tilde{\gamma})$;  

\item the specific  property of the Hadamard-Gutzwiller model is that
 all orbits $\gamma$ have the same positive Lyapunov exponent $\lambda^{(F)}$; this
 clearly does not hold in generic systems; then the action difference
$\delta S$ expressed in terms of $\varepsilon$ depends in general on $\gamma$;

\item the singularities $x_S \in \ssc$ of the map $\phi$ affect the number of self-intersections 
with small crossing angles $\varepsilon$ and may even `destroy' the partner orbit $\tilde{\gamma}$ if 
$\gamma$ approaches  $\ssc$ too closely. 
\end{itemize}  

We shall see in the following sections that working in the Poincar\'e surface of section 
enables one to resolve all these difficulties.


\section{The phase-space approach} \label{sec-phase_sp}
%
\subsection{Orbits with two almost time-reverse parts} \label{sec-TATRP}

\begin{center}
\begin{figure}
\centering
\includegraphics[width=7.3cm]{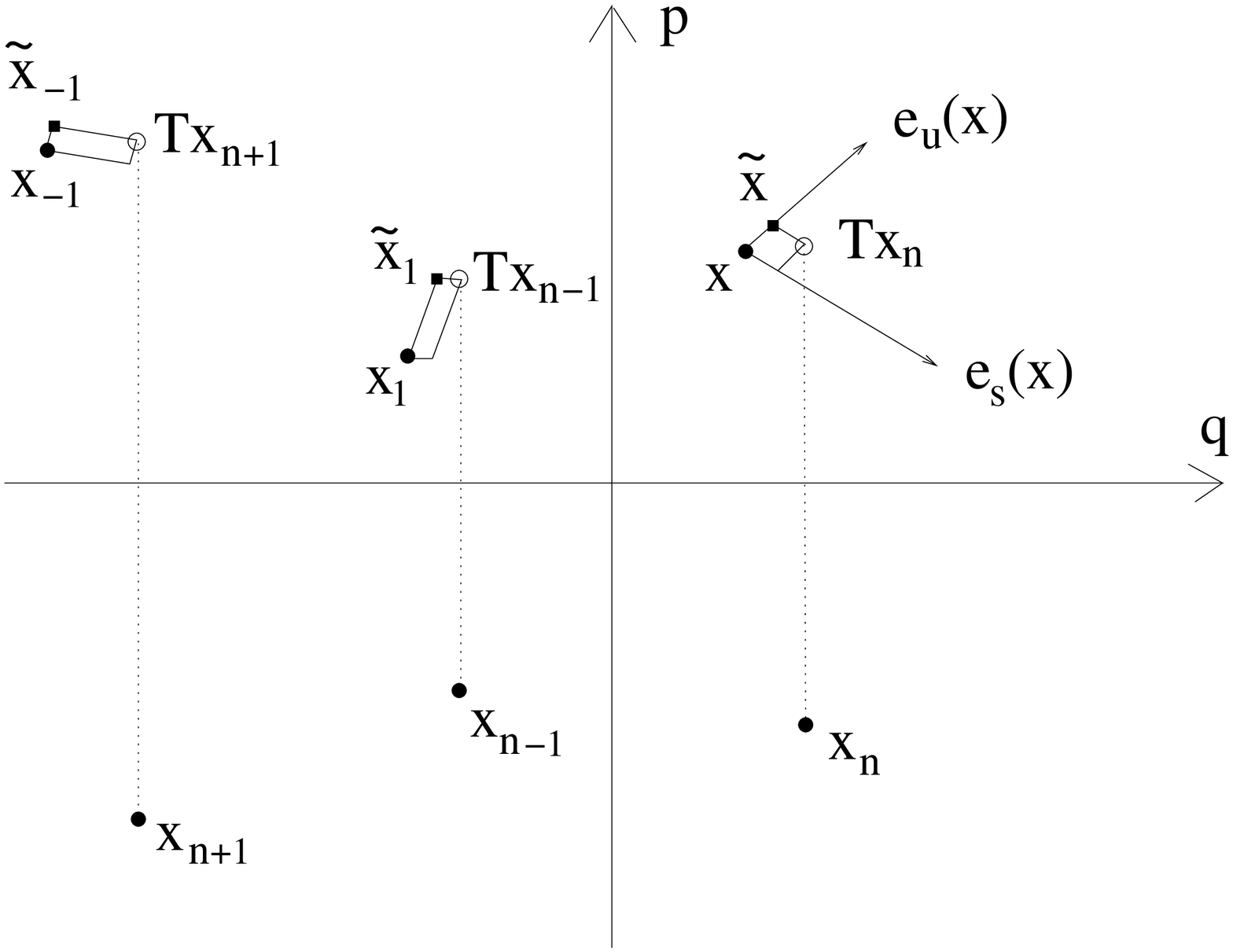}
\hspace*{0.5cm}
\includegraphics[width=7.3cm]{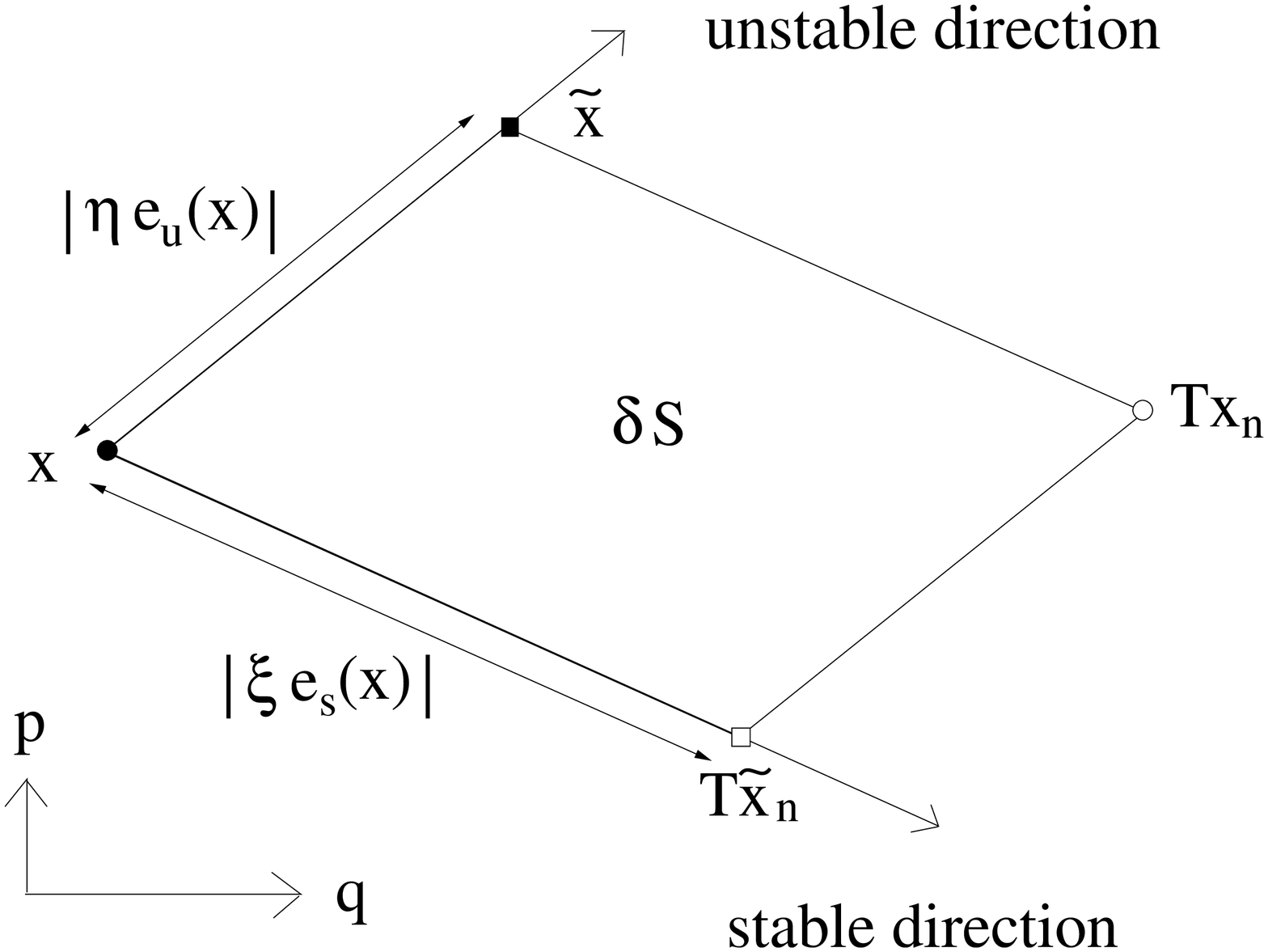}
\caption{\it (a) The two families $\{ x_m \}$ and $\{ \tilde{x}_m \}$ in the 
surface of section $\Sigma$ (only few points are represented).
(b) Magnification of (a) near $x=x_0$. 
The $N$-periodic points $x$,  $\tilde{x}$, $T x_{n}$,
and $T \tilde{x}_{n}$
 pertain to $\gamma$ (filled circles),   
$\tilde{\gamma}$ (filled squares), the TR of $\gamma$ (empty circles),
and the TR of $\tilde{\gamma}$
(empty squares).}
\label{fig-2}
\end{figure}
\end{center}

As noted in~\cite{Braun02}, if an orbit $\gamma$ has a
 self-intersection at $\qv_c$ with a small crossing angle $\varepsilon$ in configuration space,
there are  two phase-space
points $\xv_{c,i}=(\qv_c,\pv_{c,i})$ and $\xv_{c,f}=(\qv_c,\pv_{c,f})$ on $\gamma$
which are nearly TR of one another, $\xv_{c,i} \simeq T_\Gamma \,\xv_{c,f}$. Indeed, 
$| \pv_{c,i} + \pv_{c,f} | \simeq |  \pv_{c,i}| | \varepsilon|$ is very small for 
$|\varepsilon| \ll 1$
(see Fig.~\ref{fig-1}). 
There is in fact a part of $\gamma$ centred on $\xv_{c,i}$    
almost coinciding with the TR of
another part of $\gamma$, centred on $\xv_{c,f}$. 
The smaller the distance between $T_\Gamma \,\xv_{c,f}$ and $\xv_{c,i}$, the longer are
these parts of orbit.
There is therefore
a {\it family of points}  $\xv_m$ of intersection of $\gamma$ 
with the surface of section $\Sigma$, with coordinates $x_m = \phi^m x$ 
such that
\begin{equation} \label{eq-small_distances}
T x_{n-m}  \approx  x_m \;,\;m=0,\pm1, \pm 2,\ldots
\end{equation}
(see Fig.~\ref{fig-2}(a)).
The integer  $n$ is the time  (for the map) separating the two centres $x=x_0$ and $x_n$
of the two almost TR parts\footnote{
There is an analogy between the family of points $\{ \xv_m \}$ and the 
family of vertices  visited two times by an orbit in quantum graphs~\cite{Schanz02}.} 
 of $\gamma$.

It turns out that the breaking of the linear approximation (LA) plays an essential role
in the existence of a family $\{ \xv_m \}$ with property (\ref{eq-small_distances}).
Indeed, we will show that, if the orbit $\gamma$ is unstable and $n$ is large, the displacements
\begin{equation} \label{eq-Deltax_m}
\Delta x_m =  T x_{n-m} - x_m 
\end{equation}
cannot be determined  from $\Delta x = \Delta x_0$ 
by using the LA for $m \geq n/2$.
In order to make quantitative statements, and with the aim 
of transforming (\ref{eq-small_distances}) into a 
precise definition, we introduce a small real number $c_x^{(t)}$, depending on 
$x$ and on an integer $t$, the latter denoting the current time. 
Loosely speaking, $c_x^{(t)}$ is the phase-space scale
 at which deviations from the  LA  after $t$ iterations of a point $y$ near $x$ 
start becoming important. 
More precisely, this number is defined as the maximal
distance $| y_m - x_m|$ between the $m$-fold iterates of $y$ and $x$, for an arbitrary
$y \approx x$ and an arbitrary time $m$ between $0$ and $t$, 
such that the final displacement
$y_{t} - x_{t}$ can be determined from the initial displacement $y-x$ by using the LA,
$y_{t} - x_{t} \simeq M_x^{(t)} (y-x)$ 
(recall that $M_x^{(t)}$  is the linearised $t$-fold iterated map)\,\footnote{
A more quantitative definition of $c_x^{(t)}$ and
the precise meaning of `$y_{t} - x_{t} \simeq M_x^{(t)} (y-x)$'
are given in section~\ref{sec-sing} below.}.
As the errors of the LA may accumulate at each iteration, the larger the time  $t$, 
the smaller must be $c_x^{(t)}$. We shall
see below that
$c_x^{(t)}$ decreases to zero like $t^{-1}$ for large $t$ if the map $\phi$ is smooth. 
If $\phi$ is not smooth,  a typical trajectory in $\Sigma$
approaches a singularity point $x_S$ of $\phi$ arbitrarily closely between times $0$ and $t$ 
as $t \rightarrow \infty$.
As a result, $c_x^{(t)}$ decreases to zero faster than $t^{-1}$  at large $t$
($c_x^{(t)}$ even vanishes if $x$
hits $x_S$ after $m \leq t$ iterations, but such
 $x$ form a set of measure  zero).

We can now define the time $m_0$ of breakdown of the LA for the displacements 
(\ref{eq-small_distances}) as the  largest integer such that 
\begin{equation} \label{eq-familyx_m}
 | \Delta x_m | \leq c_{x}^{(m_0)}   \;\;,\;\; m=0, \ldots,m_0 \;.
\end{equation}
In other words, $m_0$ is equal to the largest integer $m$ such that 
$\Delta x_{m} \simeq M_x^{(m)} \Delta x$. 
Similarly, going backward in time, we denote by $m_0^{T}$  the largest integer $m$
such that $\Delta x_{-m} \simeq M_x^{(-m)} \Delta x$.
In what follows, we  say  that the orbit $\gamma$ has {\it two almost TR parts 
separated by $n$}
whenever (\ref{eq-familyx_m}) holds true for a family $\{ \xv_m \}$ of points
 of intersection of $\gamma$ with $\Sigma$,
where $\Delta x_m$ is defined by (\ref{eq-Deltax_m}).
The point $\xv_0$ is chosen among $\{ \xv_m  \}$ in such a way that 
$| \Delta x_0 | $ is minimum for $m=0$. 
This condition fixes $n$. In order to simplify the notation, we shall drop 
the index $0$ for the coordinate $x_0$ of the centre point $\xv_0$, writing $x=x_0$ and,
similarly, $\Delta x = \Delta x_0$.
Since we are interested in the limit $|\Delta x_0 | \ll c_x^{(m_0)}$, 
we always assume that $m_0$ and $m_0^{T}$  are large  (but much smaller than the period 
of  $\gamma$). 
If $\gamma$ is unstable, then $|\Delta x_m| \simeq | M_x^{(m)} \Delta x |$ and 
$|\Delta x_{-m}| \simeq | M_x^{(-m)}  \Delta x |$ grow
exponentially fast with $m$ for large $m$ with the same rate $\lambda_\gamma$, 
$\lambda_\gamma = \lambda_x >0$ being the 
positive Lyapunov exponent of $\gamma$ for the Poincar\'e map. 
Moreover, the components of $\Delta x$ in the
stable and unstable directions are roughly the same, since, by assumption,
$|\Delta x_m|$ is minimum for $m=0$. This implies that $m_0^{T} \approx m_0$. 

Let us first assume that $n$ is large.
The exponential growth of  $|\Delta x_m|$ in the regime of validity $m \leq m_0$ of the LA
has the following consequence.
Let us look at the  distance in configuration space in Fig.~\ref{fig-1} between 
the point $\qv_{m}$, moving on the lower branch of the right loop as one increases
$m$ (starting at $m=0$),  and its
`symmetric point' $\qv_{n-m}$, 
moving backward in time on the upper branch of the same loop.
After the time $m=n/2$, the 
two points on the lower and upper branches of the loop
are exchanged. Thus, the distance between these two points  cannot increase
for $m \geq n/2$. In contrast, it
must  decrease exponentially 
as $m$ approaches $n$, and come back to its initial value $|\qv_n - \qv_0|$ for $m=n$. 
It follows that the LA must break down before $m = n/2$, i.e., one has $m_0 \leq n/2$.
A similar reasoning holds in phase space. 
We first note that the $(n- m)$-fold iterates of $T x_{n}$ and $x$ are equal to 
$T x_m$ and $x_{n-m}$, 
respectively. Thanks to (\ref{eq-Deltax_m}),
$\Delta x_{n-m}  =  - T \Delta x_m $ for
any integer $m$.
The equality  $|\Delta x_{n-m}| = | T \Delta x_m |$ 
 would be violated if $2 m_0 \geq n \gg 1$, in view of
the exponential growth of $|\Delta x_m|$ predicted by the LA. 
As a result, for large $n$, the condition
\begin{equation}
n \geq  2 m_0
\end{equation}
must be fulfilled.

Another situation arises when  $n$ is of order $1$, $n \ll  m_0$. 
Then the above-mentioned arguments do not apply since, if $m$ is of order $1$, the  unstable and 
stable components of  $\Delta x_m$
are of the same order and  $|\Delta x_m|$ does not necessarily increase with $m$.  
Since $|\qv_{n-m} - \qv_m | \leq |T x_{n-m}-x_m| \ll 1$  
for all times $m$ between $- m_0^T$ and $m_0 \gg n$, the right 
loop in Fig.~\ref{fig-1} consists of two almost parallel lines, connected by a small piece of 
line with 
length of order $\langle l \rangle$, $\langle l \rangle$ being the mean length of a trajectory
between two consecutive intersections with $\Sigma$. 
This means that the  orbit $\gamma$ has an almost self-retracing part in $\qv$-space, 
centred at $\qv_0$.

To conclude, we have shown that
 $n_0=2 m_0$ has the meaning of a minimal  time separating two {\it distinct} 
almost TR parts of $\gamma$ (i.e., excluding almost self-retracing parts). A similar result is
obtained in~\cite{Marko,Mueller03} for continuous times. 
The continuous-time version of $n_0$ is the 
minimal  time $T_0$ to close a loop in $\qv$-space introduced in~\cite{Sieber01}.
In the present context, this time arises with the new interpretation of the breakdown of the LA.

Let $N$ be the period of $\gamma$ for the Poincar\'e map. 
If the family $\{ \xv_m \}$ fulfils condition (\ref{eq-familyx_m}), then the family 
of almost TR points $\{ \xv_{n+m} \}$  also fulfils this condition,
with $n$ replaced by $N-n$. 
This expresses the fact that, for a periodic orbit,
 the existence of a right loop in $\qv$-space implies 
the existence of a left loop (Fig.~\ref{fig-1}). 
Setting $y=x_n$, one has $T y_{N-n-m} - y_m = - T \Delta x_{-m}$ and thus
$| T y_{N-n-m} - y_m  | \leq \| T \| \,c_x^{(-m_0^T)}$ for $m=0, \ldots , m_0^T$. 
This indeed shows that
if $\{ \xv_m \}$ satisfies (\ref{eq-familyx_m}), then this is also the case for
$\{ \xv_{n+m} \}$ with $n$ replaced by $N- n$,
 $m_0$ by $m_0 ' \approx m_0^T$, and 
$m_0^{T}$ by 
${m_0^{T}}' \approx m_0$.  

The distinction made in the previous section between 
a self-intersecting orbit and an orbit with an avoided crossing in $\qv$-space
is irrelevant in the surface of section $\Sigma$: both orbits have 
two parts which are almost TR of one another. In other words, they
both have families of points $\{ \xv_m \}$ and $\{ \xv_{n+m} \}$  satisfying  
(\ref{eq-familyx_m}).
Note that these two families 
 can  correspond in $\qv$-space with   a family of  self-intersections, 
as in the case of  focusing billiards if  self-intersections occur at conjugate 
points~\cite{Mueller03,Mueller}, or with one self-intersection (or one avoided crossing) only, 
as in the case of the Hadamard-Gutzwiller model~\cite{Sieber01}.


\vspace*{-0.7cm}

\subsection{The partner orbit} \label{sec-partner_orbit} \label{eq-partner}

We can now construct 
the partner orbit $\tilde{\gamma}$ described in section~\ref{eq-RS_theory}
in the surface of section $\Sigma$.
Let $\gamma$ be an unstable orbit of period $N$ with two almost TR parts separated by 
$n < N$.
The orbit $\tilde{\gamma}$ is defined by a $N$-periodic point 
$\tilde{x} = \tilde{x}_0$ lying close to $x = x_0$. This point is such that 
\begin{equation} \label{eq-def-partner}
\begin{cases}
| T \tilde{x}_{n-t} - x_t | \ll 1  & \text{for $t=0, \ldots, n$} \\
|\tilde{x}_{t} - x_t | \ll 1 & \text{for $t=n, \ldots, N$.} 
\end{cases}
\end{equation}
It can be checked in Fig.~\ref{fig-1} that these properties indeed define the desired 
partner orbit. 
Note the symmetry of (\ref{eq-def-partner}) with respect to the exchange of $x$ and $\tilde{x}$.
By determining $\delta x = \tilde{x} - x$ as a power
series in $\Delta x$, it is shown in appendix~\ref{app-partner} that 
$x$ has at most one partner point $\tilde{x}$. 
These arguments indicate moreover that $\tilde{x}$ exists 
if $|\Delta x|$ is `sufficiently small' and the two almost TR parts of $\gamma$ are sufficiently
far apart from a singularity point of $\phi$. 
To first order in $\Delta x$, it is found in  appendix~\ref{app-partner} that
\begin{equation} \label{eq-A1}
\delta x = \tilde{x} - x 
 = T \bigl( 1 - M^{(n)}_x M^{(N-n)}_{Tx} \bigr)^{-1}  \bigl( M^{(n)}_x + T \bigr) \Delta x   
\end{equation}
in agreement with~\cite{Sieber01}. The matrices 
$M_x^{(n)}$ and $M^{(N-n)}_{Tx}$ appearing in (\ref{eq-A1}) are the stability matrices  
of the right loop and of the TR of the left loop in Fig.~\ref{fig-1}.

The partner point associated with $x_m$,  $ - m_0^{T} \leq  m  \leq m_0$,
 coincides with the $m$-fold iterate 
$\tilde{x}_m=\phi^m \tilde{x}$ of $\tilde{x}$.
This can be seen by noting that $\tilde{x}_m$ satisfies (\ref{eq-def-partner}) with $x$ 
replaced by $x_m$ and $n$ replaced by $n-2m$, as follows by combining  
(\ref{eq-def-partner}) with (\ref{eq-familyx_m}).
Hence, by uniqueness, $\tilde{x}_m$ is the partner point of $x_m$.
It is not difficult to check  this statement explicitly  to lowest order in $\Delta x$
on (\ref{eq-A1}) (see appendix~\ref{app-partner}).
We conclude that  the partner points of all points $x_m$, 
$-m_0^{T} \leq m \leq m_0$,
belong  to the same orbit $\tilde{\gamma}$.
In other words,  if  $| \Delta x | \ll 1$, there is a unique partner orbit $\tilde{\gamma}$
associated with the whole family $\{ \xv_m \}$.
If this family is  almost self-retracing, i.e., if $n \ll m_0$,
this orbit coincides with $\gamma$ itself, as already noted elsewhere~\cite{Mueller}. 
Actually, then $\tilde{x}=x$
satisfies (\ref{eq-def-partner}),
hence $\tilde{\gamma}=\gamma$ by uniqueness of the partner point
 (within the LA, this can be seen  
by replacing  $T \Delta x = - \Delta x_n = - M_x^{(n)} \Delta x$ 
in (\ref{eq-A1}); the identity $\Delta x_n = M_x^{(n)} \Delta x$ follows from 
$n \leq m_0$). 
By using a similar argument, one shows 
that the orbit  $\tilde{\gamma}'$ constructed from 
the family $\{ \xv_{n+m} \}$ is the TR of $\tilde{\gamma}$, 
as is immediately clear in Fig.~\ref{fig-1}.

\subsection{A simple example: the baker's map} \label{sec-exemple}

The main advantage of the above-mentioned construction of the pairs $(\gamma,\tilde{\gamma})$
is that it works whatever the dimension of $\Sigma$ (i.e., for systems
with $f > 2$ degrees of freedom as well). Moreover, it applies  to hyperbolic maps. 
It is instructive 
to exemplify this construction in the 
case of the baker's map. Then $\Sigma$ is the unit square.
It is convenient to equip $\Sigma$ with the distance $|x- x'| = \max \{ | q - q'|, |p-p'| \}$.
A point $x=(q,p) \in \Sigma$ is in one-to-one correspondence with a
bi-infinite sequence 
$\omega = \cdots \omega_{-2} \,\omega_{-1} \,. \,\omega_0 \,\omega_1\,\omega_2 \cdots$, obtained
from the binary decompositions
of $q$ and $p$ ($q = \sum_{l>0} \omega_{l-1}\, 2^{-l}$ and $p = \sum_{l>0} \omega_{-l}\, 2^{-l}$),
with binary symbols $\omega_l \in \{ 0, 1\}$, $l \in \integer$.
The map $\phi$ acts
on $\omega$ by shifting the point `.' one symbol to the right. The TR 
symmetry is the reflection with respect to the diagonal of the square, $T: (q,p) \mapsto (p,q)$. 
This corresponds to 
reversing the order of  the symbols of $\omega$, i.e., $T : \omega \mapsto
\omega^{T} = \cdots \omega_{2} \, \omega_1\,\omega_0 \,.\,\omega_{-1}\,\omega_{-2}\,  \cdots$. 
Periodic points are associated with sequences $\omega$ containing a finite word 
$\omega_0 \cdots \omega_{N-1}$, which repeats itself periodically; one usually
writes the finite word only, keeping in mind that circular permutations of this word correspond to
the same orbit.
It is easy to see that the condition (\ref{eq-familyx_m}) with $c_x^{(m_0)}  
= 2^{-s}$
is satisfied if $\omega_{n-l-1} = \omega_{l}$ for any $l=-s, \ldots, m_0 + s-1$.
Similarly, the condition $| \Delta x_{-m} | \leq 2^{-s}$, $m=0, \ldots, m_0^T$, is satisfied if
$\omega_{n-l-1} = \omega_{l}$ for any $l=-m_0^T -s, \ldots, s-1$.
This means that $\omega$ has the form
\begin{equation}
x \;\;\longleftrightarrow\;\;
 \omega =  Z^{T}_{L} \, L  \, Z_{L}\,.\, Z_{R} \, R \, Z^{T}_{R} \;,
\end{equation}
where  $Z_{L}$, $Z_{R}$, $L$, and $R$ are finite words containing 
$(m_0^{T}+s)$, $(m_0+s)$,  $(N- 2 m_0^{T}-2s)$, and $(n- 2 m_0-2s)$ 
symbols, respectively. 
The symbol sequence
of the partner point $\tilde{x}$ is obtained  by reversing time on $R$ 
and leaving all other symbols unchanged,
\begin{equation}
\tilde{x}\;\;  \longleftrightarrow\;\; 
\tilde{\omega} = Z^{T}_{L}\,  L\, Z_{L}\,.\, Z_{R} R^{T} Z^{T}_{R} \;.
\end{equation}
The inequality $n > 2 m_0+ 2s$ must be 
fulfilled in order that $R$ is nonempty. In the opposite case,
$\omega$ has an almost self-retracing part 
$Z_{L} Z_{R} Z^{T}_{R} Z^{T}_{L}$ and $\tilde{\omega} = \omega$.
Similar pairs $(\omega, \tilde{\omega})$ of symbol sequences occur in the 
Hadamard-Gutzwiller model~\cite{Heusler02} and in certain billiards~\cite{Mueller}. 
The families $\{ x_m \}$ and $\{ \tilde{x}_m \}$ look like those in Fig.~\ref{fig-2}(a) 
after a rotation by an angle $\pi/4$.

\section{Use of the unstable and stable coordinates} \label{sec-unstable_comp}

To evaluate the leading off-diagonal correction $K_2(\tau)$ to the form factor,
we shall first consider the second sum in (\ref{eq-K2}) over all partner orbits $\tilde{\gamma}$
of $\gamma$,
for a fixed unstable  periodic orbit $\gamma$, which will be assumed to be infinitely long and to cover 
densely and  uniformly the surface of  section.
We will then argue in section~\ref{sec-form_fact} that one can 
replace the obtained result inside the sum over $\gamma$ 
in the limit $T \rightarrow \infty$.
The sum over the partner orbits $\tilde{\gamma}$ of $\gamma$ is to be expressed as an integral over 
some continuous parameters characterising 
$\tilde{\gamma}$, chosen such that the action difference 
$\delta S= S_{\tilde{\gamma}} - S_\gamma$  
is a function of these parameters only. In configuration space, one may
integrate over the crossing angle 
$\varepsilon$~\cite{Sieber01}. 
It is argued  in this section that a
convenient choice  of  parameters in the surface of section is given by   
the unstable and stable coordinates of the small displacement $\Delta x$.
The local coordinate system defined by the  unstable and stable
directions is singled out by the stretching and squeezing properties of the 
dynamics. These properties play a crucial role  
in the theory of Sieber and Richter, because they determine the  time $m_0$
of breakdown of the LA and the exponential smallness of the distances 
(\ref{eq-def-partner}).

\subsection{The coordinate family $\lc_{x,\eta,\xi}$}

Under the hyperbolicity assumption (ii), there are at almost all $y \in \Sigma$
two vectors 
$e_{u} (y)$ and $e_{s}(y)$ tangent to the unstable and stable manifolds at $y$, 
which span the whole tangent space $\tc_y \Sigma$.  
These vectors  can be found by means of
the cocycle decomposition~\cite{Gaspard}, 
\begin{equation} \label{eq-e_u,s}
M_y^{(m)} e_{u} (y)   =   \Lambda_{u,y}^{(m)} \, e_{u} ( y_m ) \;\;\mbox{ , }\;\;
M_y^{(m)} e_{s} (y)   =   \Lambda_{s,y}^{(m)}   \, e_{s} ( y_m )
\;,
\end{equation}
where $\Lambda_{u,y}^{(m)}$ and $\Lambda_{s,y}^{(m)}$ are the stretching 
and squeezing factors.
Because $M_y^{(m)}$ is symplectic,  
$\Lambda_{s,y}^{(m)} = 1/\Lambda_{u,y}^{(m)}$ and  the symplectic product 
$e_{u}(y) \wedge e_s(y)$ is independent of $y$ (see~\cite{Gaspard}). 
The  vectors $e_{u,s}(y)$ can be `normalised' in such a way that
this constant is equal to $1$,
\begin{equation} \label{eq-symplectic_prod_e_us}
e_u(y) \wedge e_s (y) = 1\;.
\end{equation}
The product of the norms of $e_{u}(y_{m})$ and $e_{s} (y_m)$    
diverges  as $m \rightarrow \pm \infty$ if the angle between the unstable and stable 
directions at $y_m$ decreases to zero. 
Since the exponential growth of $M_y^{(m)} e_{u} (y)$ at large $m$
is (by definition) captured by the stretching factor,
the divergence of $|e_{u,s} (y_m)|$ is smaller than exponential, 
$\ln |e_{u,s}(y_m)| = o( m)$~\cite{Gaspard}.
The notation $f(m) = o (m)$, where  $f$ is an arbitrary  function over integers, stands for  
$f(m)/m \rightarrow 0$ as $m \rightarrow \pm \infty$.
The stretching factor 
$\Lambda_{y}^{(m)} = \Lambda_{u,y}^{(m)}$ satisfy
\begin{equation} \label{eq-Lyapunov}
\ln | \Lambda^{(m)}_y | = m \lambda_y + o(m) 
\end{equation}
where  $\lambda_y$ is the positive Lyapunov exponent at $y$.
If $y$ belongs to a periodic orbit $\gamma$ with period $N$, then $e_{u,s}(y)$ are the eigenvectors
of the stability matrix $M_y^{(N)}$ of $\gamma$ and $| \Lambda_y^{(N)}| = \exp ( N \lambda_\gamma )$.
By invoking the TR symmetry, $M_{T y}^{(m)} T = T  M_{y}^{(-m)}$.
Replacing this expression into (\ref{eq-e_u,s}), one finds 
$ e_{u,s} (Ty) \propto T e_{s,u} (y)$,
with some $\phi$-invariant proportionality factors. By ergodicity, these factors 
are almost everywhere constant ($y$-independent). One can thus `normalise' 
$e_{u,s}(Ty)$ in such a way that   
\begin{equation} \label{eq-TR_e_us}
  e_{u}(Ty) = T  e_{s} (y)  \;\mbox{ , }\;  e_{s} (Ty) = T  e_{u} (y)   \;\mbox{ , }\;
\Lambda_{Ty}^{(m)} = \Lambda_{y_{-m}}^{(m)} 
\end{equation}
for almost all $y \in \Sigma$.
Note that this agrees with (\ref{eq-symplectic_prod_e_us}), since
$ T  e_{s}(y) \wedge T e_u( y) = e_{u}(y) \wedge e_{s} (y)$.

\begin{center}
\begin{figure}
\centering
\includegraphics[width=7.3cm]{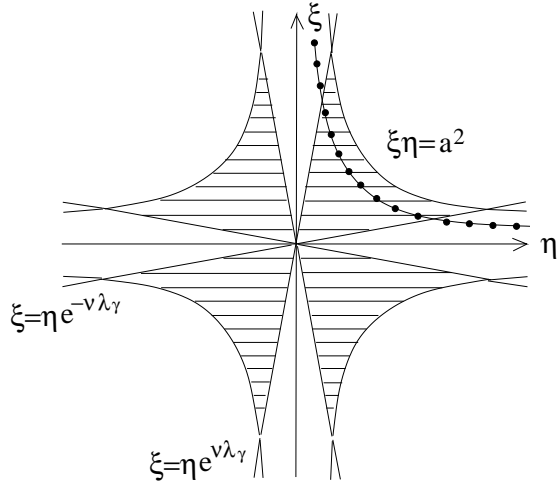}
\caption{\it Domain $\dc_{\gamma,a,\nu}$ in the $(\eta,\xi)$-plane 
(region marked by horizontal lines). 
The black points are the points $(\eta_m,\xi_m)$ in $\lc_{x,\eta,\xi}$;
$\nu$ of them are contained in  $\dc_{\gamma,a,\nu}$.}
\label{fig-4}
\end{figure}
\end{center}

Two almost TR parts of an unstable orbit $\gamma$ 
can be parametrised by the family
\begin{equation} 
{\mathcal{L}}_{x,\eta,\xi} 
 = \left\{ (\eta_m,\xi_m) \,; \,- m_0^{T} \leq m  \leq m_0 \right\}
\subset \real^2
\end{equation}
 of the unstable and stable coordinates
$(\eta_m,\xi_m)$ of the displacements $\Delta x_m$,
\begin{equation} \label{eq-xfamilies2}
\Delta x_m  =  T x_{n-m} - x_m = \eta_m \, e_{u}(x_m) + \xi_m\, e_{s} (x_m) 
\;.
\end{equation}
Thanks to (\ref{eq-e_u,s}), 
\begin{equation} \label{eq-eta_m}
\eta_m = \Lambda^{(m)}_x\, \eta \;\mbox{ , }\; \xi_m = \frac{\xi}{\Lambda^{(m)}_x}
\;\mbox{ , }\;\eta_m\,\xi_m = \eta\,\xi \;,
\end{equation}
where $- m_0^{T} \leq m \leq m_0$ and $\eta = \eta_0$, $\xi = \xi_0$.
The points $(\eta_m,\xi_m) \in \lc_{x,\eta,\xi}$ are located on a hyperbole 
in the $(\eta,\xi)$-plane (see Fig.~\ref{fig-4}).

In the case of the baker's map (section~\ref{sec-exemple}), 
$e_{u}(y)$ and $e_s(y)$ are independent of $y$ and coincide with the unit vectors in the 
$q$- and $p$-directions. The stretching factors $\Lambda_y^{(m)}=2^m$ are also $y$-independent. 
The coordinates $\eta_m$ and $\xi_m$ are the usual $q$- and $p$-coordinates of 
$\Delta x_m = T x_{n-m} - x_m$,
\begin{equation} \label{eq-baker}
\left\{
\begin{array}{ccccc}
\eta_m & =  &  p_{n-m} - q_m & = & 2^m ( p_n - q)\\
 \xi_m & =  &  q_{n-m} - p_m & = & 2^{-m} (q_n - p) \,.
\end{array}
\right.
\end{equation}
%


\vspace*{-0.7cm}

\subsection{Estimation of $m_0$} \label{sec-m_0}

In the limit $\eta,\xi \rightarrow 0$, the time $m_0$ of  breakdown
 of the LA depends logarithmically on the unstable coordinate $\eta$,
\begin{equation} \label{eq-meta}
m_0 = - \frac{\ln |\eta|}{\lambda_\gamma} 
          + o \left( \lambda_\gamma^{-1}  \ln |\eta| \right) \;.
\end{equation} 
Indeed, thanks to hyperbolicity, 
$|\Delta x_m|$ grows exponentially fast with $m$ with the rate $\lambda_\gamma>0$,
until it reaches, for $m=m_0$, a value of the order of the phase-space scale
$c_x^{(m_0)}$ at which deviations from the  LA start becoming important.   
Since
$|\Delta x_{m_0} | \approx | \eta_{m_0}| $,
one must have $\ln (| \eta|/c_x^{(m_0)} ) \sim - m_0  \,\lambda_\gamma$.   
More precisely, we may approximate $|\Delta x_{m} |$ by $|\eta_{m}| |e_{u}(x_{m})|$
for $m=m_0$ and $m=m_0+1$, making an exponentially small error for large $m_0$
(recall that $|\eta| |e_{u} (x)| \approx |\xi| |e_s(x)|$).
By definition,
$|\Delta x_{m_0} |$ is smaller than $c_{x}^{(m_0)}$ and 
$|\Delta x_{m_0+1} |$  is greater than 
$c_{x}^{(m_0+1)}$. Then (\ref{eq-meta}) follows from
(\ref{eq-Lyapunov}), (\ref{eq-eta_m}), and 
$\ln | e_{u}(x_{m}) | =  o (m)$.
Note that the terms 
 $\lambda_\gamma^{-1} |\ln c_x^{(m_0)}|$ and $\lambda_\gamma^{-1} |\ln c_x^{(m_0+1)}|$ 
have been neglected in  (\ref{eq-meta}). As stated in section~\ref{sec-TATRP},
for a smooth map $\phi$,  
$c_x^{(m_0)}$ decreases to zero like $m_0^{-1}$ as $m_0 \rightarrow \infty$, i.e., as
$\eta \rightarrow 0$. Therefore, $|\ln c_x^{(m_0)}|$ is of order 
$\ln m_0 = \ooc (\ln | \ln |\eta| |)$ and can be incorporated in the error 
term in (\ref{eq-meta}). If $\phi$ has singularities on $\Sigma$, 
 $c_x^{(m_0)}$ decreases to zero faster than $m_0^{-1}$ as $\eta \rightarrow 0$.
In such case, it will be argued in section~\ref{sec-sing}
that formula (\ref{eq-meta}) is not valid for {\it all} orbits $\gamma$. 
However, the right-hand side of (\ref{eq-meta}) always gives an upper
bound on $m_0$.
Strictly speaking, the asymptotic behaviours (\ref{eq-Lyapunov}) and
(\ref{eq-meta}) provide  good approximations 
only if $m_0$ is close to a multiple of (or is much larger than) the period $N$ of $\gamma$. 
The physically relevant values of $m_0$ are, however, such that $1 \ll m_0 \ll N$.
For such $m_0$'s, (\ref{eq-meta})  should
give nevertheless a reasonable approximation
of the average value of $m_0$ (in fact it  gives a good approximation
of the inverse of the average of the inverse of $m_0$). This average can be taken  
over all points $\xv$  on $\gamma \cap \Sigma$ 
satisfying  (\ref{eq-familyx_m})  such that the unstable and stable
coordinates of $\Delta x = T x_n - x$ are in small intervals $[\eta,\eta+ \D \eta]$
and $[ \xi, \xi+ \D \xi]$, for an arbitrary integer $n \leq N/2$ and some fixed $\eta$, $\xi$,
$\D \eta \ll |\eta| \ll 1$,  and $\D \xi \ll | \xi | \ll 1$.

One shows similarly that
\begin{equation} \label{eq-mxi}
m_0^{T} =  - \frac{\ln |\xi|}{\lambda_\gamma}  + o ( \lambda_\gamma^{-1} \ln | \xi | ) \;.
\end{equation}
For the baker's map, in view of (\ref{eq-baker}),
$2^{m_0} | p_n - q | \leq 2^{-s} \leq 2^{m_0+1} | p_n -q|$, where we have chosen
$c_x^{(m_0)} = 2^{-s}$. If $m_0 \gg s$, this yields $m_0 \simeq - \ln | p_n - q| /\ln 2$  
and, similarly, $m_0^T \simeq - \ln | q_n - p| /\ln 2$,
in agreement with (\ref{eq-meta}) and (\ref{eq-mxi}).

\subsection{The probability of `near-head-on return'} \label{sec-P}

To count the number of partner orbits of an orbit $\gamma$ with a very large period $N$,
one needs to know the probability to have two points 
on $\gamma$ which are nearly TR of one another. 
The aim of this subsection is to determine the (unnormalised) 
probability density $P_{\gamma}(\eta,\xi)$   
associated with the unstable and stable coordinates of 
$\Delta x_t = T x_{t+n} - x_t$, for all pairs $(\xv_t, \xv_{t+n})$ of almost TR points on 
$\gamma \cap \Sigma$ which do not pertain to an almost self-retracing family 
(i.e., such that $n \geq 2 m_0$). This density is defined through 
the number $P_{\gamma}(\eta,\xi) \,\D \eta \, \D \xi$   of points $\xv_t$ on $\gamma \cap \Sigma$
 such that 
the unstable and stable coordinates of 
$\Delta x_t$ are in the 
infinitesimal intervals $[\eta,\eta + \D \eta]$ and $[ \xi, \xi + \D \xi]$, for an arbitrary 
integer $n$ between $2 m_0 (x_t,\eta)$ and $N/2$.
Let us recall that  the partner orbits $\tilde{\gamma}$
and $\tilde{\gamma}'$ built from the  two families $\{ \xv_{m}  \}$ and  
$\{ \xv_{n+m} \}$, separated from their almost TR families by the times  $n$ and $N-n$, respectively, 
are TR of one another (section~\ref{sec-phase_sp}). The two pairs 
$(\gamma,\tilde{\gamma})$ and $(\gamma, \tilde{\gamma}' )$ have thus
identical contributions to the form factor (\ref{eq-K2}) (the corresponding action differences
$\delta S$ are clearly the same). 
This is why it suffices to consider only the  
orbits $\tilde{\gamma}$ constructed from the family
with the smaller time, $n \leq N/2$.

Let us define the infinitesimal parallelograms $\D \Sigma_{x, \eta, \xi}$ in $\Sigma$
by
\begin{equation} \label{eq-DV}
\D \Sigma_{x, \eta, \xi} = \Bigl\{ y \in \Sigma \; ; \;
\eta  \leq  ( y  - x )_{u} \leq   \eta + \D \eta \; ,\;
\xi  \leq   ( y - x )_{s}  \leq   \xi + \D \xi \;
\Bigr\} 
\end{equation}
where $(y-x)_{u,s}$ 
are the unstable and stable coordinates of
$y-x$.
Then
\begin{equation} \label{eq-1}
{P}_{\gamma} (\eta, \xi) \,\D \eta\,\D \xi 
 =  
  \sum_{t=0}^{N -1} 
   \sum_{n= 2 m_0(x_t,\eta)}^{N/2} \chi \bigl( T x_{n+t} \in \D \Sigma_{x_t,\eta,\xi} \bigr)  
\end{equation}
where $\chi(\pc)$ equals $1$ if the property $\pc$ is true and $0$ otherwise.
We shall assume here that the periodic orbit $\gamma$
covers densely and uniformly the surface of section $\Sigma$.
If $N \gg 4 m_0 \gg 1$, the sum over $n$
can then be replaced by a phase-space integral, giving
\begin{equation} \label{eq-2}
{P}_{\gamma}^{\,\text{erg}}   (\eta, \xi)\,\D \eta\,\D \xi
 =  
   \sum_{t=0}^{N -1} 
     \left( \frac{N}{2}  - 2 m_0(x_t,\eta)  \right) 
      \int  \D \mu (y) \,\chi \bigl( T y \in \D \Sigma_{x_t,\eta,\xi} \bigr) 
\;. 
\end{equation}
This integral is nothing but the area
$|T \D \Sigma_{x_t,\eta,\xi}| = | \D \Sigma_{x_t,\eta,\xi}|$
of  the parallelogram (\ref{eq-DV})
per unit area. By (\ref{eq-symplectic_prod_e_us}), it is equal to 
$\D \eta \,\D \xi/|\Sigma|$. 
In virtue of (\ref{eq-meta}),
\begin{equation} \label{eq-DP_gamma}
{P}_{\gamma}^{\,\text{erg}}   (\eta, \xi)
 \simeq 
\frac{N}{2 |\Sigma|} \left( N   + \frac{4\,\ln | \eta|  }{\lambda_\gamma} 
  \right)\;.
\end{equation}
It is worth 
noting that the  ergodic hypothesis implies the identity 
between (\ref{eq-1}) and (\ref{eq-2}) for a set of points $x_0$ of measure one, and
does not tell  anything {\it a priori} about the points $x_0$ on periodic orbits, 
of  measure zero in $\Sigma$.
We shall argue below that, although (\ref{eq-1}) and (\ref{eq-2}) may differ for 
individual periodic orbits $\gamma$ which do not cover $\Sigma$ uniformly, 
one can use the ergodic result  (\ref{eq-DP_gamma})
to calculate the form factor in the semiclassical limit.

\subsection{The domain $\dc_{\gamma,a,\nu}$} \label{sec_L}

The density $P_\gamma(\eta,\xi)$ just defined overcounts the number
of partner orbits $\tilde{\gamma}$ relevant for the form factor. 
Actually,  a unique partner orbit $\tilde{\gamma}$ is associated
with each family $\{ \xv_m; -m_0^T \leq m \leq m_0  \}$ (section~\ref{eq-partner}),
whereas all points $\xv_m$ belonging to the same family are 
counted separately in $P_\gamma(\eta,\xi)$.
To avoid  overcounting, we define a domain $\dc_{\gamma,a,\nu}$
in the  $(\eta,\xi)$-plane $\real^2$, having the good property to contain,
for any $(\eta,\xi)$ inside this domain, a {\it fixed} 
number $\nu$ of 
elements $(\eta_m,\xi_m)$ in the family
${\mathcal{L}}_{x,\eta,\xi}$. This integer $\nu$ is independent of $\eta$ and $\xi$ (and 
thus of $m_0$ and $m_0^T$) and is such that $1 \ll \nu \ll N$.
Provided that this condition is fulfilled, 
the precise value of $\nu$ does not matter for the final result.
Introducing also a small number $a>0$ controlling the maximal values of $|\eta|$ and $|\xi|$,
we define
\begin{equation} \label{eq-criterium}
\dc_{\gamma,a,\nu} = \left\{ (\eta,\xi )  \in \real^2 \mbox{ ; } 
e^{-\nu \lambda_\gamma} \leq \frac{|\xi |}{|\eta |} \leq   
 e^{\nu \lambda_\gamma}
\mbox{ , }  |\eta\, \xi | \leq a^2 \; \right\} \;.
\end{equation}
If  $(\eta,\xi)$ belongs to $\dc_{\gamma,a,\nu}$, then
$|\eta|$ and $|\xi|$ are bounded by 
$a \, e^{\nu \lambda_\gamma/2}$. 
The domain $\dc_{\gamma,a,\nu}$ is represented in Fig.~\ref{fig-4}.
For any $(\eta,\xi) \in \dc_{\gamma,a,\nu}$, it contains  
$d_x \simeq \nu$ elements of the family ${\mathcal{L}}_{x,\eta,\xi}$.
Actually, in view of (\ref{eq-Lyapunov}) and (\ref{eq-eta_m}),   
\begin{equation} \label{eq-as_behav}
\ln \left( \frac{|\xi_m|}{|\eta_m|} \right) 
 = \ln \left( \frac{|\xi|}{|\eta|} \right)
  - 2   m \lambda_\gamma  + o (m ) \;\mbox{ , } \; 1 \ll | m | \leq 
 \min\{ m_0, m_0^{T} \} \;.
\end{equation}
By choosing $a$ small enough,
one has, thanks to (\ref{eq-meta}), 
$m_0 + o (m_0) \geq \nu$ for any $(\eta,\xi) \in \dc_{\gamma,a,\nu}$
(for instance, if $c_x^{(m_0)} = b\, m_0^{-\alpha}$ with $b,\alpha >0$, 
one may choose $a = b\,e^{- 3 \nu \lambda_\gamma/2}$).
The number of $(\eta_m,\xi_m)$ in the family $\lc_{x,\eta,\xi}$
which fulfil the first condition  in (\ref{eq-criterium}) is then equal to $\nu + o (\nu)$.
If $(\eta,\xi) \in  \dc_{\gamma,a,\nu}$, 
the second condition $|\eta_m\, \xi_m | \leq a$  is fulfilled, by (\ref{eq-eta_m}), for all $m$ 
between $-m_0^{T}$ and $m_0$, since it holds true for $m=0$. 
Hence  $\dc_{\gamma,a,\nu}  \cap \lc_{x,\eta,\xi}$ has
$d_{x} = \nu + o (\nu)$ elements.
Note that, as already stressed in section~\ref{sec-m_0},
 the use of the asymptotic behaviour (\ref{eq-as_behav}) for $1 \ll m_0 \ll N$ is in fact only
justified if one is concerned with the average value of 
$d_{x}^{-1}$, taken e.g. over all $x$ on $\gamma$ 
satisfying  (\ref{eq-familyx_m}) with unstable and stable
coordinates of $\Delta x$  in some intervals $[\eta,\eta+ \D \eta]$
and $[ \xi, \xi+ \D \xi]$ for an arbitrary $n \leq N/2$.
 
Let us define a new  weighted probability density $\tilde{P}_\gamma (\eta,\xi)$, in
which  the overcounting of partner orbits is compensated by
a weight 
$d_{x_t}^{-1}$ attributed to each event 
$T x_{n+t} \in  \D \Sigma_{x_t,\eta,\xi}$ in (\ref{eq-1}). 
By repeating the argument of the last subsection, one gets
\begin{equation} \label{eq-3}
\tilde{P}_\gamma^{\,\text{erg}} (\eta,\xi) 
 =  
   \sum_{t=0}^{N -1} 
     \frac{1}{d_{x_t}} 
      \left( \frac{N}{2}  - 2 m_0(x_t,\eta)  \right) \frac{1}{|\Sigma|}
 \simeq  
\frac{N}{2 |\Sigma| \nu} \left( N   + \frac{4\,\ln | \eta|  }{\lambda_\gamma} 
  \right) \;.
\end{equation}
This density differs from (\ref{eq-DP_gamma}) by a factor $1/\nu$.

\subsection{Action difference}
\label{sec-action_diff}

The main point in determining the action difference  
$ \delta S = S_{\tilde{\gamma}} - S_\gamma$ of the two orbits 
 $\tilde{\gamma}$ and $\gamma$ 
is to observe the  following geometrical property of the partner points in the small 
$|\Delta x|$  limit: 
$x$, $\tilde{x}$, $T x_n$ and $T \tilde{x}_n$  form a parallelogram,
with sides parallel to $e_{u,s}(x)$ (see Fig.~\ref{fig-2}(b)).
It may be tempting to argue that, since, by (\ref{eq-def-partner}),
$\tilde{x}$ must be exponentially close to the 
unstable manifold at $x$ and the stable manifold at $T x_{n}$, 
this property follows 
straightforwardly from the continuity of the unstable and stable directions. However,
some care must be taken here. Indeed, the unstable and stable directions
vary notably inside the small region between  
the four $N$-periodic points $x$, $\tilde{x}$, $T x_n$, and $T \tilde{x}_n$. 
This is due to the well-known intricate pattern built by the unstable 
and stable manifolds in the vicinity
of heteroclinic points.   
 We proceed as follows. Since $T x_n - x = \eta\, e_u(x) + \xi\, e_s(x)$, it suffices to show that, 
to lowest order in $\Delta x$,
\begin{equation} \label{eq-deltax}
 \tilde{x} - x   =  \eta\,e_{u} (x)  
\;\mbox{ , }\;
  T \tilde{x}_n - x   =  \xi \,e_{s} (x) \;.
\end{equation}   
The idea is to combine a stability analysis 
with the fact that $e_{u}(x)$ is nearly proportional to $e_u(\tilde{x})$.
For  indeed, the orbits 
$\gamma$ and $\tilde{\gamma}$ look almost the same between times
$t=-(N-n)$ and $t=0$. Therefore, their unstable directions must be almost parallel 
at $x$ and $\tilde{x}$. 
Similarly,  the TR  of $\gamma$
is very close to $\tilde{\gamma}$ between $t=0$ and $t=n$, so that the stable directions
at $T x_n$ and $\tilde{x}$ must be almost parallel,
$e_s (  T x_n  ) \propto e_s(\tilde{x} )$.

To show (\ref{eq-deltax}), let us consider 
 the unstable and stable coordinates $(\psi,\zeta)$ 
 of $x - \tilde{x}$,
\begin{equation} \label{eq-psi_zeta}
x - \tilde{x} =  \psi\, e_{u}(\tilde{x}) + \zeta  \, e_{s}(\tilde{x}) \;.
\end{equation}
In view of (\ref{eq-def-partner}), 
one may approximate 
$M_x^{(n)}$ by $M_{T \tilde{x}_n}^{(n)}$ and $M_{Tx}^{(N-n)}$ by $M_{T \tilde{x}}^{(N-n)}$
if $|\Delta x| \ll 1$.
By (\ref{eq-A1}), one has, to lowest  order in $\Delta x$,
\begin{equation} \label{eq-linear_deltax}
- \bigl(  1 - M_{T \tilde{x}}^{(N)} \bigr) T ( x - \tilde{x} ) =  (M_x^{(n)} + T ) \Delta x\;.
\end{equation} 
Here
$M_{T \tilde{x}}^{(N)} = M_{T \tilde{x}_n}^{(n)} M_{T \tilde{x} }^{(N-n)}$ is
the stability matrix 
of the TR of $\tilde{\gamma}$, with eigenvectors
$e_{u,s} (T \tilde{x})$ and eigenvalues 
$\Lambda_{\tilde{\gamma}}^{\pm 1}$ such that
$| \Lambda_{\tilde{\gamma}}|= \exp  (N \lambda_{\tilde{\gamma}})$.
By using (\ref{eq-TR_e_us}), (\ref{eq-xfamilies2}), and
(\ref{eq-psi_zeta}) and by neglecting terms smaller by a factor   
$\exp  ( - N \lambda_{\tilde{\gamma}})$ or $\exp (-n\lambda_\gamma)$ than the other terms, 
(\ref{eq-linear_deltax}) can be rewritten as
\begin{equation}
 \zeta \, \Lambda_{\tilde{\gamma}} \, e_u ( T \tilde{x})
   - \psi \, e_s ( T \tilde{x}) 
 = \Lambda_x^{(n)} \eta \, e_{u} ( x_n )  
   + \eta\,e_{s} (Tx)\;.
\end{equation} 
Hence, for $n \gg 1$ and $(N-n) \gg 1$,   $\zeta  \simeq 0$ and $\psi \, e_{s}(T \tilde{x}) \simeq 
- \eta \, e_s (Tx)$. 
Replacing this result into (\ref{eq-psi_zeta}) and using (\ref{eq-TR_e_us}), we arrive at the 
first equality in (\ref{eq-deltax}).
We now argue that the partner point of $Tx$ is $\tilde{x}_n$. 
This is already clear in
Fig.~\ref{fig-1}. This can be shown by invoking the uniqueness of the partner point and 
 by noting that the replacement of $(x, \tilde{x})$ by $(Tx, \tilde{x}_n)$ and of
$n$ by $N-n$  in (\ref{eq-def-partner}) leads to the exchange of the
upper and lower lines, up to a TR. This  replacement  gives, by (\ref{eq-Deltax_m}),  
$\Delta (Tx) = T \Delta x = \xi\,e_{u} ( T x) + \eta\,e_{s} (Tx)$.
Then the second identity in (\ref{eq-deltax}) is a consequence of the 
first one (with the above-mentioned replacement), to which one applies the TR map $T$.

The action difference $\delta S$
is determined to lowest order in $\Delta x$ 
in appendix~\ref{app-actiondiff}. It coincides with the symplectic area 
of the parallelogram $(x,\tilde{x},T x_n,T \tilde{x}_n)$,
\begin{equation} \label{eq-deltaScont2}
\delta S 
 =  
 ( \tilde{x} - x ) \wedge ( T \tilde{x}_n  - x ) 
    =  \eta \, \xi  \;, 
\end{equation} 
where we have chosen $L P$ as the unit of action. 
It is clear that $\delta S$ is independent of the choice of the pair of partner 
points $(x_m,\tilde{x}_m)$, with $-m_0^{T} \leq m \leq m_0$, as 
all these pairs $(x_m,\tilde{x}_m)$ correspond to the same orbit pair $(\gamma,\tilde{\gamma})$. 
Since 
$\eta_m \,\xi_m$ is the only $m$-independent combination of $\eta_m$ and $\xi_m$ of second order, 
the result (\ref{eq-deltaScont2}) (with an unknown prefactor) was thus to be expected.

\section{Leading off-diagonal correction to the form factor} \label{sec-form_fact}

\subsection{The case of smooth maps} \label{sec-form_factor_smooth}

The form factor (\ref{eq-K2}) is, introducing a dimensionless
Planck constant $\heff = \hbar/(L P)$, 
\begin{equation} \label{eq-K_2}
K_2 (\tau) 
 =  \frac{2}{T_H} \, 
  \frac{1}{\delta T} 
   \left\langle \;
\sum_{T \leq  T_\gamma \leq T+\delta T} 
A_\gamma^2 
   \int_{\dc_{\gamma,a,\nu} } \D \eta \,\D \xi\,
    \tilde{P}_{\gamma} (\eta,\xi) \,
    \exp \left( \frac{\I \,\eta \,\xi}{\heff} \right) 
   \right\rangle_E \, .
\end{equation}
The variables $\eta$ and $\xi$ are integrated over the domain 
$\dc_{\gamma,a,\nu}$ defined in  (\ref{eq-criterium}). 
As seen above,  to avoid overcounting the partner orbits,
one must use the weighted
density $\tilde{P}_\gamma(\eta,\xi)$, related to the density $P_\gamma(\eta,\xi)$
defined in section~\ref{sec-P} by a factor $1/\nu$.
Only partner orbits constructed from parts of $\gamma$
separated by $n \leq N/2$ from their almost TR parts are taken into account in these
near-head-on-return densities, where $N$ is the period of $\gamma$ for the map $\phi$.
The other partner orbits, corresponding to $n \geq N/2$, 
 give the same contribution to the form factor (see section~\ref{sec-P}). 
This contribution is taken into account by the factor $2$ in (\ref{eq-K_2}).

The values of $\eta$ and $\xi$ contributing significantly to the integral (\ref{eq-K_2})
are  of order $\sqrt{\heff}$.
Thanks to (\ref{eq-meta}), $n_0=2 m_0$ is thus of the order of the Ehrenfest time 
$\lambda^{-1}_\gamma | \ln \heff|$. For large periods, one has
$N  \simeq T /\langle t_y \rangle = \tau |\Sigma|/(2 \pi \heff)$,  where 
\begin{equation}
\langle t_y \rangle = \int \,\D \mu(y) \,t_y 
= ( |\Sigma| L P )^{-1} \int  \D \yv \,\delta \bigl( H(\yv) - E \bigr) 
= \frac{(2 \pi \hbar)^2 \,\overline{\rho}(E)}{|\Sigma| L P}
\end{equation}
is 
the mean first-return time.  Therefore 
$N  \gg n_0 \gg 1$ for the physically relevant values of $\eta$ in the semiclassical limit.
This has also the important consequence that, for small but finite $\heff$, 
the values of the time $T =  \tau\,T_H$ for which the theory of Sieber and Richter
works are limited below by the Ehrenfest time $2 T_0 \simeq 2 \langle t_y \rangle n_0$,
since $N$ must be bigger than $2 n_0$.

We would  now like to replace $\tilde{P}_\gamma (\eta,\xi)$  
by the ergodic result (\ref{eq-3})
inside the sum (\ref{eq-K_2}). 
To do this, one needs that  
long periodic orbits  are  uniformly  distributed  in phase space, 
in the sense explained in section~\ref{sec-hyp} (see also~\cite{HODA84}). 
We shall  {\it assume} here that this is the case, and  that 
(\ref{eq-3}) can indeed be used 
under the sum  over periodic orbits  (\ref{eq-K_2})  in the limit $T \rightarrow \infty$. 
A good indication supporting this assumption
is given by Bowen's equidistribution theorem~\cite{Bowen72}: for any continuous function $f$ on $\Gamma$, 
\begin{equation} \label{eq-eq_theo}
\sum_{T \leq T_\gamma \leq T + \delta T} 
 e^{-\lambda_\gamma^{(F)}\, T_\gamma} \int_0^{T_\gamma} \D t \,
f( \xv_\gamma(t) ) 
 \sim 
   \sum_{T \leq T_\gamma \leq T + \delta T} T_\gamma \,  
     e^{-\lambda_\gamma^{(F)} \, T_\gamma} 
    \int_\Gamma \D \mu_E (\yv) \, f (\yv )
\end{equation}
as $T \rightarrow \infty$.  The integral on the left-hand side
is taken along $\gamma$, and $\lambda_\gamma^{(F)}$ is the positive 
Lyapunov exponent of $\gamma$ for the Hamiltonian flow. 
The normalised microcanonical measure 
$\D \mu_E(\yv ) =  \nc \, \delta ( H(\yv) - E ) \, \D \yv$  
on the right-hand side 
is the product of the invariant measure $\mu$ and the Lebesgue measure 
along the orbit~\cite{Gaspard}, 
\begin{equation} \label{eq-productmeasure}
\int_\Gamma \D \mu_E(\yv) \,f(\yv ) 
 = 
  \int_\Sigma \D \mu (y) \,F(y) 
\;\;\;,\;\;
F(y) \equiv \frac{1}{\langle t_y \rangle} \, \int_0^{t_y} \D t \,f(\yv(t) ) \;.
\end{equation}
Orbits $\gamma$ with multiple traversals
$r_\gamma \geq 2$  have a negligible contribution in (\ref{eq-eq_theo})
because they are exponentially less numerous than the orbits with 
$r_\gamma=1$.
One can thus replace $T_\gamma^2 \,\exp ( -\lambda_\gamma^{(F)} \,T_\gamma )$
by  the square amplitude $A_\gamma^2$ in 
 (\ref{eq-eq_theo}), 
\begin{equation} \label{eq-dist_theo_PS}
\sum_{T \leq T_\gamma \leq T + \delta T} A_\gamma^2 
 \sum_{n=0}^{N -1} F(x_n) 
  \sim 
   \sum_{T \leq T_\gamma \leq T + \delta T} N \,A_\gamma^2 \int \D \mu(y) \, F(y)
\;\; \;, \;\;\;T \rightarrow \infty \;.
\end{equation}
To our knowledge, the sum rule (\ref{eq-eq_theo}) has been proved rigorously 
for a restricted class of systems only, which includes 
uniformly hyperbolic systems~\cite{Bowen72} and the free motion on a Riemann
surface with non-negative curvature~\cite{Knieper98}.
Moreover, 
(\ref{eq-dist_theo_PS}) cannot be applied directly 
to our problem, because  $\chi$ and $m_0$ in (\ref{eq-1}) are discontinuous functions.
We shall not  pursue here in trying to motivate the above-mentioned assumption. 
Instead, we shall
 go ahead in determining $K_2(\tau)$. It would be interesting from a 
mathematical point of view to find general conditions on the dynamics implying 
our assumption.

Replacing $\tilde{P}_\gamma (\eta,\xi)$ by (\ref{eq-3}) into the integral
\begin{equation}
I_{\gamma,a,\nu} = 
\int_{\dc_{\gamma,a,\nu}} \D \eta \,\D \xi \,
   \tilde{P}_{\gamma} (\eta,\xi) \,\exp \left( 
    \frac{\I  \,\eta \,\xi}{\heff} \right) \;,
\end{equation}
one obtains
\begin{eqnarray} \label{eq-K22}
\nonumber
I_{\gamma,a,\nu} 
&  = & 
\frac{2 N \,\heff}{|\Sigma| \nu} 
\left\{ 
    \int_{0}^{a\,e^{-\nu \lambda_\gamma/2} } \frac{\D \eta}{\eta} 
     \left( N   + \frac{4\ln \eta }{\lambda_\gamma}  \right)  
       \sin \left( \frac{\eta^2 \,e^{\nu \lambda_\gamma} }{\heff} \right)
 \right.
\\
\nonumber
&  &
\left.
   + \int_{a\,e^{-\nu \lambda_\gamma/2} }^{a\,e^{\nu \lambda_\gamma/2} }  \frac{\D \eta}{\eta} 
  \left( N   + \frac{4 \ln \eta }{\lambda_\gamma} \right)
    \sin \left( \frac{a^2}{\heff} \right)
\right.
  \\
&  &
\left.
   - \int_{0}^{a\,e^{\nu \lambda_\gamma/2} } \frac{\D \eta}{\eta} 
    \left( N   + \frac{4  \ln \eta}{\lambda_\gamma} \right) 
    \sin \left( \frac{\eta^2 \,e^{-\nu \lambda_\gamma} }{\heff }\right) 
\right\}  
\;. 
\end{eqnarray}
The first and third integrals
can be computed with the help of the changes of variables $\eta' = \eta \,e^{\nu\lambda_\gamma/2}$
and $\eta' = \eta \,e^{-\nu\lambda_\gamma/2}$, respectively. This yields
\begin{equation}
I_{\gamma,a,\nu}
 =  \frac{T}{\pi\,T_H}
  \left\{  - 4 \int_0^{a} \frac{\D \eta'}{\eta'}
     \sin \left( \frac{\eta'^2}{\heff} \right)
       + \lambda_\gamma \left( N   + \frac{4 \ln a}{\lambda_\gamma}  \right) 
       \sin \left( \frac{a^2 }{\heff} \right)
   \right\}
\;.
\end{equation}
The first term inside the brackets is equal to $-\pi + \ooc ( \heff\, a^{-2} )$.
The second one is a rapidly oscillating sine and
gives rise to higher-order contributions in $\hbar_{\text{eff}}$  after
the energy average.
Ignoring this oscillating term and the terms of order $\heff/a^2$, 
one gets $I_{\gamma,a,\nu} = -T/T_H$. It should be stressed that this 
result is  true only  for very long
periodic orbits which cover  uniformly  the whole surface of section $\Sigma$.
It has been argued above that, although  such a result is not true for all orbits $\gamma$, it can
be used {\it inside} the sum over $\gamma$ in (\ref{eq-K_2}). This gives
\begin{equation}
K_2 (\tau) 
 = - \frac{2 T}{T_H^2} \, \frac{1}{\delta T} 
  \sum_{T \leq  T_\gamma \leq T+\delta T} 
   A_\gamma^2 \Bigl( 1 + \ooc \bigl( \heff \,a^{-2} \bigr) 
 \Bigr)
\;.
\end{equation}
We can now invoke the 
Hannay-Ozorio de Almeida sum rule~\cite{HODA84},
\begin{equation}
\frac{1}{\delta T} \sum_{ T \leq T_\gamma \leq T+ \delta T} A_\gamma^2 \sim T
\;\;,\;\; T \rightarrow \infty\;,
\end{equation}
to arrive at the announced result
\begin{equation} \label{eq-result}
K_2 (\tau) = - 2 \tau^2 \;,
\end{equation}
valid  in the limit
$\hbar \rightarrow 0$, $\tau = T /(2 \pi \hbar \,\overline{\rho}(E) )$ fixed.

\subsection{The case of maps with singularities} \label{sec-sing}

We have ignored so far the fact that $\phi$ or its derivatives may be 
singular on a closed set $\ssc \subset \Sigma $ of measure zero, as is typically the case 
in billiards~\cite{Katok}.
As stressed above,
 the term $\lambda_\gamma^{-1}\,|\ln c_{x}^{(m_0)} |$ neglected in (\ref{eq-meta})
can be as large as $m_0$ 
if $\gamma$ approaches $\ssc$ too closely between times $0$ and $m_0$.
In such a case, it may {\it a priori} also happen that no partner orbit is associated with the 
family $\{ \xv_m \}$ (see appendix~\ref{app-partner}).
The aim of this section is to show that, under assumptions (iv) and (v) of section \ref{sec-hyp},
the result (\ref{eq-result}) is still valid.
Indeed, we shall see that (\ref{eq-meta}) and the 
action difference (\ref{eq-deltaScont2})
are correct for all $x$ outside a small subset of $\Sigma$. This subset turns out to
be  unimportant for 
$K_2(\tau)$ in view of  its  negligible measure. 
We will not discuss here the diffractive corrections to the semiclassical 
expression (\ref{eq-K_semicl}), which should  
 {\it a priori} also be taken into account.

Let us first estimate the phase-space scale $c_x^{(t)}$ associated with the breakdown of the LA 
introduced in section~\ref{sec-TATRP}.
By invoking the cocycle property 
$M_x^{(t)} = M_{x_{t-1}}^{(1)} \ldots M_{x_{1}}^{(1)} M_{x_{0}}^{(1)}$ of the linearised map,
it is easy to show that $y_t - x_t $ is equal to
\begin{equation} \label{eq-Taylor}
M^{(t)}_x (y-x) +  \frac{1}{2}  \sum_{m=0}^{t-1} \sum_{\alpha,\beta=1}^2 M_{x_{m+1}}^{(t-1-m)}  
  \left( \frac{\partial^2 \phi}{\partial x^\alpha \partial x^\beta}  \right)_{x_m} 
    \left[ M_x^{(m)} (y-x) \right]^\beta
     \left[  M_x^{(m)} (y-x)\right] ^{\alpha} + \cdots \;,
\end{equation}
where $M_x^{(0)}$ is the identity matrix.
The displacement
$y_t - x_t$ can be determined from the initial displacement $y-x$ by using the LA 
if the first term
of the Taylor expansion (\ref{eq-Taylor})
is much greater than the subsequent (higher-order) terms.
This is the case if $|y_m - x_m | \leq c_{x}^{(t)}$ for $0 \leq m \leq t$, with
\begin{equation} \label{eq-c_k}
c_{x}^{(t)} 
 = 
  \frac{{b}}{t} \min_{m=0,\ldots,t-1}  
   \min_{r \geq 2} \min_{\alpha_1,\ldots,\alpha_r=1,2} 
   \left|
    \left( \frac{\partial^r \phi}{\partial x^{\alpha_1} \ldots \partial x^{\alpha_r}} \right)_{x_m} 
   \right|^{-1/(r-1)} 
    \;.
\end{equation}
A small fixed  number  ${b} \ll 1$ controlling the error of the LA
has been introduced. 
By assumption~(v) of section~\ref{sec-hyp}, 
\begin{equation} \label{eq-maj_c_x}
c_x^{(t)} \geq \frac{{b}}{t} \,C_2^{-1}\, \min_{m=0,\ldots,t-1} 
 d ( x_m, \ssc)^{\sigma_2}  \;.
\end{equation}

Let $\delta >0$ and
\begin{equation}
B_{\delta,S}^{(m_0)} 
 = 
  \bigcup_{m=0}^{m_0-1} \phi^{-m} ( B_{\delta,S}) \;\;\;,\;\;\;
B_{\delta,S} = \Bigl\{ x \in \Sigma; | x - x_S | \leq \delta\;\,{\text{for some}}\;\, x_S \in \ssc \Bigr\}\;. 
\end{equation}
By~(iv), it is possible to choose $\delta$ such that the probability 
to find $x$ in $B_{\delta,S}^{(m_0)}$,
\begin{equation}
\mu \left( B_{\delta,S}^{(m_0)} \right) \leq \sum_{m=0}^{m_0-1} \mu ( B_{\delta,S}) 
 \leq C_1\,m_0 \,\delta^{\sigma_1}
\end{equation}
is very small. For instance,
taking $\delta = ({b}/m_0 )^{1/\sigma_1}$ gives 
$\mu (B_{\delta,S}^{(m_0)}) \leq C_1 {b} \ll 1$.
Let us assume that $x$ is not in
$B_{\delta,S}^{(m_0)}$, i.e., that the part of orbit  between times $0$ and $m_0-1$ 
does not approach a singularity
closer than by a distance $\delta$.
Then, by  (\ref{eq-maj_c_x}), 
\begin{equation}
c_{x}^{(m_0)}  \geq   C_2^{-1} \, {b}^\sigma  \,m_0^{-\sigma}
\;,
\end{equation}
with $\sigma = 1+\sigma_2/\sigma_1$. 
Therefore, 
$| \ln c_{x}^{(m_0)} |$ is at most of order
$\ln m_0 = \ooc ( \ln  |\ln |\eta| | )$ 
as $\eta \rightarrow 0$ and can be incorporated in the error term in (\ref{eq-meta}).
This reasoning shows that (\ref{eq-meta}) can be used except 
if the centre point $x$ of the family $\{ x_m \}$ is in  
$B_{\delta,S}^{(m_0)}$.

If $x$ is in $B_{\delta,S}^{(m_0)}$, then $m_0$ may have a different 
behaviour for $|\eta| \ll 1$ than that  
 given by (\ref{eq-meta}). Anomalous behaviours due to singularities 
of the minimal time $T_0$ to close a loop
have been indeed observed in numerical simulations for the desymmetrized diamond billiard
and the cardioid billiard in~\cite{Mueller03,Mueller}. These numerical results show that
non-periodic orbits satisfy  (\ref{eq-meta}), with $\lambda_\gamma$ replaced
by the mean positive Lyapunov exponent $\langle \lambda \rangle$, except those orbits
approaching too closely a singularity.

By using an expansion similar to (\ref{eq-Taylor}),  
one can show that 
the relative errors made by approximating
 $M_x^{(n)}$ by $M_{T \tilde{x}_n}^{(n)}$ and $M_{Tx}^{(N-n)}$ by $M_{T \tilde{x}}^{(N-n)}$
are small in the small $| \Delta x|$ limit if $x$ is not in
$B_{\delta,S}^{(m_0)}$. The arguments of
section~\ref{sec-action_diff} leading to the parallelogram $(x,\tilde{x},Tx_n,T\tilde{x}_n)$
and to the 
action difference (\ref{eq-deltaScont2}) thus apply if  $x$ is not in
$B_{\delta,S}^{(m_0)}$.

Let us now parallel the calculation of  $\tilde{P}_{\gamma}(\eta,\xi)$ of 
sections~\ref{sec-P} and \ref{sec_L}.
Replacing the time average over $t$  in (\ref{eq-3}) by a phase-space average,
\begin{eqnarray} \label{eq-DP_x}
\tilde{P}_{\gamma}^{\,\text{erg}} (\eta, \xi)  
& = &  
  N   \int   \D \mu (x)  \frac{1}{d_x} 
  \left( \frac{N}{2} - 2 m_0(x,\eta) \right) \frac{1}{|\Sigma|} 
\\
\nonumber
&  = & 
  \frac{N}{2 | \Sigma| \nu}
 \left\{
   N - 4 \,\overline{m}_0 
   \Bigl( 1 - \mu \bigl( B_{\delta,S}^{(\overline{m}_0 )} \bigr) \Bigr) 
     - 4  \int_{B_{\delta,S}^{(\overline{m}_0 )} } \D \mu (x) \, m_0 (x,\eta) 
    + o \bigl( \overline{m}_0 \bigr)
  \right\}\;,
\end{eqnarray}
with $\overline{m}_0  = - \lambda_\gamma^{-1} \ln |\eta|$ and 
$\delta = (b/\overline{m}_0 )^{1/\sigma_1}$.
The integral in the second line gives a negligible contribution, as
$m_0(x) \leq \overline{m}_0  + o ( \overline{m}_0 )$ 
for any $x$ (section~\ref{sec-unstable_comp}) and 
$\mu (B_{\delta,S}^{(\overline{m}_0  )}) \leq C_1 {b}$.
Thus  $\tilde{P}_{\gamma}^{\,\text{erg}} (\eta,\xi)$ is still given by (\ref{eq-3}), with an error
of order ${b}$. As ${b}$ can be chosen arbitrarily small (in the limit
$\hbar \rightarrow 0$), it follows that  $K_2 (\tau) = -2 \tau^2$
for Poincar\'e maps with singularities satisfying hypotheses (iv) and (v)
 of section~\ref{sec-hyp}.

\section{Conclusion}

We have proposed a new method  to calculate the contribution of the
Sieber-Richter pairs of periodic orbits  
to the semiclassical form factor in chaotic systems 
with TR symmetry. Our basic assumption is the hyperbolicity of the classical dynamics.  
The method has been illustrated for 
Hamiltonian systems with two degrees of freedom.
By assuming furthermore that
long periodic orbits are uniformly distributed in phase space, 
 the same leading off-diagonal correction $K_2(\tau) = - 2 \tau^2$ 
as found in~\cite{Sieber01}
for the Hadamard-Gutzwiller model has been obtained.
This result is system independent and coincides  with the GOE prediction to second order 
in the rescaled time $\tau$. 
One  advantage of our method is its applicability to hyperbolic area-preserving  maps,
 provided their  invariant ergodic measure  
is the Lebesgue measure. This should allow one to treat the case of periodically driven systems.
Moreover, the method is suitable to
treat hyperbolic systems with more than two degrees of freedom $f$, 
for which the relevant periodic
orbits do not in general have self-intersections in configuration space.  
A Sieber-Richter pair of orbits $(\gamma,\tilde{\gamma})$ is then parametrised by $f-1$ unstable 
and $f-1$ stable coordinates $(\eta^{(1)},\ldots, \eta^{(f-1)})$
and $(\xi^{(1)},\ldots, \xi^{(f-1)})$. The  time  $m_0$ of breakdown of the linear approximation
 is given by the minimum of  
$-\ln |\eta^{(i)}|/\lambda_\gamma^{(i)}$ over all $i=1,\ldots ,f-1$,
where $\lambda_\gamma^{(i)}$ is the $i$th positive Lyapunov exponent of $\gamma$. 
For Hamiltonian systems, the action 
difference $\delta S = S_{\tilde{\gamma}} - S_\gamma$  
is given by the sum $\sum_{i} \eta^{(i)} \xi^{(i)}$.
It is $\gamma$-independent, whereas $\delta S$ depends on the stability
exponents of $\gamma$ in the approach of Sieber and Richter~\cite{Sieber01}.  
The evaluation
of the integral (\ref{eq-K_2})  is more involved for $f>2$ than for $f=2$
and will be the subject of future work.
A second advantage of the phase-space approach is that it is canonically invariant and thus
immediately applicable to systems with non-conventional time-reversal symmetries.
A third advantage is, in our opinion, that 
orbits with crossings and avoided crossings in configuration
space are treated here on equal footing.

A further understanding of the universality of spectral fluctuations in classically chaotic systems
may be gained by studying the contributions of the correlations between orbits 
with {\it several} pairs of almost time-reverse parts (`multi-loop orbits')
and their associated `higher-order' partners. These contributions are expected to
be of higher order in $\tau$.
A first step in this direction has been done recently for quantum graphs~\cite{Berkolaiko02}.
The phase-space approach presented in this work might be useful to tackle this problem. 
One would like to know if the RMT result (\ref{eq-K_GOE}) can be reproduced  in the semiclassical limit 
{\it to all orders in $\tau$} by looking at correlations between  these partner orbits  only, or if 
other types of correlations must be taken into account. 
An alternative way to study this problem is to investigate the 
impact of the partner orbits
 on the weighted action correlation function defined and studied in~\cite{Argaman93,Smilansky02}. 
 
The periodic-orbit correlations discussed in this work have also remarkable consequences for 
transport in mesoscopic devices in the ballistic regime:  they lead to weak-localization 
corrections to the conductance
in agreement with RMT~\cite{Richter02}. 
More generally, they should 
be of importance in any $n$-point correlation function of a
clean chaotic system with time-reversal symmetry.

\vspace{0.5cm}

\noindent \textbf{Acknowledgements:}
I am grateful to  S. M\"uller for explaining his work on billiards,
and  to F.~Haake and to one of the referees for their suggestions to improve  the 
presentation of the results.
I also thank P.~Braun and  S.~Heusler for enlightening discussions and
comments on  this manuscript, 
J.~Bellissard and M.~Porter for their remarks on a first version of the paper,
and H.~Schulz-Baldes and W.~Wang for interesting discussions. 

\vspace{0.5cm}

{\it Note added.} - When this work was mostly completed, I was informed that M. Turek and 
K.~Richter were working on a similar approach. Drafts of the two papers were 
exchanged during
the Minerva Meeting of Young Researchers in Dresden from January 29 to February 2, 2003.

\vspace{0.3cm}

\begin{appendix}

\renewcommand{\theequation}{\Alph{section}\arabic{equation}}

\section{Existence and uniqueness of a partner orbit} \label{app-partner}

\setcounter{equation}{0}

We present in this appendix a general method, based on a Taylor expansion, 
to prove the existence and the uniqueness 
of the partner orbit. 

Let $\gamma$ be an  orbit of period $N$ with two almost TR parts
separated by $n < N$. Let $x$ be the centre point of the family $\{ x_m; - m_0^{T} \leq m \leq m_0 \}$.
The partner orbit $\tilde{\gamma}$ is defined by an $N$-periodic  point $\tilde{x}$ in the vicinity of 
$x$, called the partner point of $x$. This point fulfils property (\ref{eq-def-partner}), i.e.,
it is such that (i) $|T \tilde{x}_{n-t} - x_t| \ll 1$ 
between times $t= 0$ and $t=n$, and (ii)~$| (T \tilde{x})_t - (T x)_t | \ll 1$ between $t=0$ and $t=N-n$.
The small displacement $\delta x = \tilde{x} - x$ 
is obtained as a power series in $\Delta x = T x_n - x$,
\begin{equation} \label{eq-ansatz2}
\delta x^\alpha =  ( \tilde{x} - x )^\alpha 
 = \sum_{r=1}^\infty 
  \bigl[ A^{(n,r)}_x \bigr]_{\beta_1 \cdots \beta_r}^\alpha 
   \Delta x^{\beta_1} \cdots \Delta x^{\beta_r}\;.
\end{equation}
We  use here the summation convention for the 
Greek indices $\beta_1,\ldots,\beta_r=1,2$, referring to the $q$- and $p$-coordinates 
in $\Sigma$ ($x^1 = q$, $x^2 = p$). Let us stress that
it is necessary to go beyond  the linear approximation (term $r=1$ in the series (\ref{eq-ansatz2})) 
to establish the existence of the partner orbit.
Indeed, one must show that $\tilde{x}$ is exactly  $N$-periodic, i.e., that 
$\tilde{x}_N = \tilde{x}$ to all orders in $\Delta x$.

Let us assume that the map $\phi$ is smooth along $\gamma$ and its TR. 
We get the coefficients $A^{(n,r)}_x$ in (\ref{eq-ansatz2})
by expanding  the final 
displacements  as Taylor series in the initial ones for (i) 
the part of $\gamma$ between $t=0$ and $t=n$, and (ii) the part of the TR of $\gamma$ 
between $t=0$ and $t=N-n$. The identity
$\tilde{x}_N = \tilde{x}$ is then used to match the two results.
More precisely, 
the computation is performed in four steps: 
(1)~expand $( T \tilde{x})_{N-n} -  ( T x)_{N-n}$ in powers of $T \tilde{x} - T x$;
(2)~replace $\delta x$ by (\ref{eq-ansatz2}) into this result;
(3)~expand  $T \tilde{x}  - x_n$ in powers of $T \tilde{x}_n  - x$ and
replace  the series obtained in the previous step  into this expansion and
(4)~identify each power of $\Delta x$. 
These manipulations lead for the linear order $r=1$ to
\begin{equation} \label{eq-A1bis}
D_x^{(n)}  T A^{(n,1)}_x 
 = M^{(n)}_x + T    
\end{equation}
with $D_x^{(n)} =  1 - M^{(n)}_x M^{(N-n)}_{Tx}$.
Let us
denote the partial derivatives
$( \partial^r \phi^t /\partial x^{\beta_1} \ldots \partial x^{\beta_r} )^\alpha$
 by $[ M_x^{(t,r)} \bigr]^\alpha_{\beta_1 \ldots \beta_r}$, 
with $t, r \geq 1$.
For any $2 \times 2$ matrix $C$, we set
$[C  A^{(n,r)}_x ]^\rho_{\beta_1 \cdots ,\beta_r} 
= C^\rho_\alpha   [A^{(n,r)}_x ]^\alpha_{\beta_1 \cdots \beta_r}$.
The higher-order tensors $A^{(n,r)}_x$, $r \geq 2$, are obtained recursively through the formula
\begin{eqnarray} \label{eq-recursiveA}
\nonumber
  \bigl[ D_x^{(n)}
    T A^{(n,r)}_x \bigr]^\rho_{\beta_1 \cdots \beta_r} 
& = & 
  \sum_{s=2}^r  \bigl[ B_x^{(n,s)} \bigr]^{\rho}_{\alpha_1 \cdots \alpha_s}
\sum_{r_1 + \cdots + r_s = r, r_i \geq 1} 
 \bigl[ T A^{(n,r_1)}_x \bigr]^{\alpha_1}_{\beta_1 \cdots \beta_{r_1} } \cdots\\
& &
 \times \bigl[ T A^{(n,r_s)}_x \bigr]^{\alpha_s}_{\beta_{r-r_s+1} \cdots \beta_r} \;,
\end{eqnarray}
with
\begin{eqnarray}
\nonumber
 \bigl[ B_x^{(n,s)} \bigr]^{\rho}_{\alpha_1 \cdots \alpha_s}
& = &
\sum_{l=1}^s  \sum_{s_1 + \cdots + s_l = s, s_i \geq 1}
  \frac{1}{l! s_1 ! \cdots s_l !}
   \bigl[ M^{(n,l)}_x \bigr]^\rho_{\delta_1 \cdots \delta_l} 
    \bigl[ M^{(N-n,s_1)}_{Tx} \bigr]^{\delta_1}_{\alpha_1 \cdots \alpha_{s_1}} 
\cdots \\
& &
 \times \bigl[ M^{(N-n,s_l)}_{Tx} \bigr]^{\delta_l}_{\alpha_{s-s_l+1} \cdots \alpha_s}  \;.
\end{eqnarray}
We have assumed for simplicity that the TR map $T$ on $\Sigma$ is linear.

It is worth noting that all tensors  $A^{(n,r)}_x$ are obtained  by 
inverting the same matrix $D_x^{(n)}$.
If $\det D_x^{(n)} \not= 0$, then (\ref{eq-A1bis})
 reduces to (\ref{eq-A1}) and all 
$A^{(n,r)}_x$'s are uniquely defined. 
 Since $1- D_x^{(n)}$ tends to the stability matrix
$M^{(N)}_{x_n} = M^{(n)}_x M^{(N-n)}_{x_n}$ of the unstable orbit $\gamma$ as 
$|\Delta x | \rightarrow 0$, 
$\det D_x^{(n)} \not= 0$ for sufficiently small 
$|\Delta x|$. This argument, however,  does not suffice to show that 
$D_x^{(n)}$ is invertible for the physically relevant values of $|\Delta x|$, which are of
order $\sqrt{ 2 \pi \heff} = \sqrt{  \tau | \Sigma|/N}$ (section~\ref{sec-form_factor_smooth}).
Another open mathematical problem concerns the convergence of the series (\ref{eq-ansatz2}).
It can be expected that (\ref{eq-ansatz2})
diverges when the orbit $\gamma$ approaches too closely a singularity
$x_S \in \ssc$ between times $-m_0^{T}$ and $m_0$.  
Provided that
$D_x^{(n)}$ is invertible  
and the series (\ref{eq-ansatz2}) converges, the $N$-periodic point $\tilde{x}$ 
 exists and is unique.

The above-mentioned  construction is not restricted to 
the centre point $x$ in the family $\{ x_m ; - m_0^T \leq m \leq m_0 \}$.
Taking another point $x_m$  in this family, one can as well construct its
partner point $\widetilde{(x_m)}$, 
by replacing $x$ by $x_m$, $n$ by $(n-2m)$, and $\Delta x$ by $\Delta x_m$
in (\ref{eq-ansatz2}). 
Let us show that, 
to linear order in $\Delta x$, $\widetilde{(x_m)}$ is
the $m$-fold iterate  $\tilde{x}_m$ of $\tilde{x}$.
To lowest order in $\Delta x$, one finds
\begin{equation}
\widetilde{(x_m)} - {x}_m 
 =   A_{x_m}^{(n-2m,1)} M_x^{(m)} \Delta x 
 =  M_{\tilde{x}}^{(m)} A_{x}^{(n,1)} \Delta x 
   =  \tilde{x}_m - x_m\;.
\end{equation}
The second equality is obtained by approximating  $1- D_x^{(n)}$ and $M^{(m)}_{x_n}$
 by  $M_{T \tilde{x}}^{(N)}$ 
and $M^{(m)}_{T \tilde{x}}$, respectively
(see section~\ref{sec-action_diff}), by
using the cocycle property of the linearised maps, and by invoking 
the TR symmetry, which implies $M_{T y}^{(m)} T = T ( M_{y_{-m}}^{(m)} )^{-1}$.
It follows that $\widetilde{(x_m)} = \tilde{x}_m$ belongs to the 
same partner orbit $\tilde{\gamma}$ as $\tilde{x}$.

To conclude, we have given strong arguments in support of the existence of a unique partner
orbit $\tilde{\gamma}$ associated with the family $\{ x_m; - m_0^T \leq m \leq m_0 \}$
if $|\Delta x| \ll 1$ and the points in this family do not approach too closely
 a singularity
of $\phi$.

\section{Action difference}
 \label{app-actiondiff}

\setcounter{equation}{0}

The action difference  
$ \delta S = S_{\tilde{\gamma}} - S_\gamma$ between the two partner orbits
 $\tilde{\gamma}$ and  $\gamma$ can be computed
by considering separately the contributions 
$\delta S_{R}$ and $\delta S_{L}$ of the right  loop (part of $\gamma$ between $\qv_0$ 
and $\qv_n$)  
and of the left loop  (part between $\qv_n$ and $\qv_N$)
in Fig.~\ref{fig-1}. $\delta S_{R}$ and $\delta S_L$ can be evaluated by means of the formula
\begin{equation} \label{eq-Sdiffgen}
S (\tilde{\qv}_i, \tilde{\qv}_f , E ) - S ( \qv_i, \qv_f , E )
 = 
\bigl( \pv_f   + \frac{1}{2} \, \delta \pv_f  \bigr)  \cdot
  \delta \qv_f 
  - \bigl( \pv_i  + \frac{1}{2} \, \delta \pv_i  \bigr) \cdot 
    \delta \qv_i + \ooc ( |\delta \xv (t)|^3 )\;,
\end{equation}
which gives the  difference of action
of two nearby trajectories $\qv(t)$ and $\tilde{\qv}(t) = \qv(t) + \delta \qv(t)$ 
of energy $E$, initial positions $\qv_i \not= \tilde{\qv}_i$
and final positions $\qv_f \not= \tilde{\qv}_f$.
This formula is  also valid for billiards. It
is easily obtained by expanding the action difference up to second order 
in $\delta \qv_i$ and $\delta \qv_f$, and by
using $\partial S/\partial \qv_f = \pv_f$ and
 $\partial S/\partial \qv_i = - \pv_i$. 
In billiards,
the momenta on the two trajectories have jumps  
$\mbox{\boldmath{$\sigma$}}  = \pv^{(+)}_{\text{refl}} - \pv^{(-)}_{\text{refl}}$ 
and $\tilde{\mbox{\boldmath{$\sigma$}}} 
 = \tilde{\pv}^{(+)}_{\text{refl}} - \tilde{\pv}^{(-)}_{\text{refl}}$
  at each reflection on the 
boundary $\partial \Omega$
(the sign $-/+$ refers to the values just prior/after the reflection).
At first glance, a new term
$
\delta S_{\text{refl}}
  = - ( \tilde{\mbox{\boldmath{$\sigma$}}}
     + {\mbox{\boldmath{$\sigma$}}} ) \cdot \delta \qv_{\text{refl}}/2
$
should then be added to (\ref{eq-Sdiffgen}) for each reflection point $\qv_{\text{refl}}$ 
on the unperturbed trajectory (such an additional term
arises when writing the action difference  as a sum of   
two contributions, corresponding
to the two segments between $\qv_i$ and 
$\qv_{\text{refl}}$ and  between  $\qv_{\text{refl}}$
and $\qv_f$).    
However, $\delta S_{\text{refl}}$ is of order $|\delta \xv (t)|^3$. Actually,
\begin{equation}
\delta \qv_{\text{refl}} = \delta q \, \Tv  
+ \frac{\delta q^2}{2}  \, \kappa \,  \Nv  + \ooc(\delta q^3 )
= \delta q \, \tilde{\Tv}   
- \frac{ \delta q^2}{2}  \, \kappa  \,  \Nv    + \ooc(\delta q^3)
\;,
\end{equation}
where $\delta q$ is the arc length on $\partial \Omega$
between the two nearby reflection points $\qv_{\text{refl}}$ and $\tilde{\qv}_{\text{refl}}$, 
$\kappa$ is the curvature and $\Tv$, $\Nv$ are the unit vectors 
tangent and normal to $\partial \Omega$ at  $\qv_{\text{refl}}$ (see Fig.~\ref{fig-0}). 
The
tangent vector $\tilde{\Tv}   = \Tv  + \delta q\, \kappa  \,  \Nv  + \ooc (\delta q^2)$ 
at $\tilde{\qv}_{\text{refl}}$ 
appears in the last expression. 
Invoking the fact that  $\mbox{\boldmath{$\sigma$}}$ and
$\tilde{\mbox{\boldmath{$\sigma$}}}$
are perpendicular to the boundary, 
one gets $\delta S_{\text{refl}} = \ooc (\delta q^3)$.

Let us denote by $\xv$, $\tilde{\xv}$, $\xv_{n}$ 
and $\tilde{\xv}_{n}$ the points on the surface of section $\Sigma$ 
with respective  $(q,p)$-coordinates 
$x,\tilde{x}, x_n$ and $\tilde{x}_n$. 
In the case of a billiard $\Omega$, these points are by definition associated  with the values
of the momenta just after a reflection on $\partial \Omega$. The corresponding points 
just before a reflection are denoted by the same letters with
an added upper subscript $(-)$. 
The momentum jumps are denoted by
 $\mbox{\boldmath{$\sigma$}}= \pv - \pv^{(-)}$, with corresponding notation for
$\tilde{\pv}$, $\pv_n$ and $\tilde{\pv}_n$. 
The action differences $\delta S_R$ and $\delta S_L$ are obtained by 
applying~(\ref{eq-Sdiffgen}) with
$$
\begin{array}{ccccccccccccccc} 
\xv_i & = & T_\Gamma \,\xv_{n}^{(-)}  & , &   \tilde{\xv}_i & = & \tilde{\xv}  &, & 
\xv_f &= & T_\Gamma\, \xv & , & \tilde{\xv}_f & = &\tilde{\xv}_{n}^{(-)} 
\\ 
\xv_i & = & \xv_{n}  & , & \tilde{\xv}_i & = & \tilde{\xv}_{n} & , & 
\xv_f & = & \xv^{(-)} & , & \tilde{\xv}_f & = &\tilde{\xv}^{(-)} \;,
\end{array}
$$
respectively. This  yields
\begin{equation}
2 \, \delta S_{R}
 = 
  - (  2 \pv  - \tilde{\pv}_{n}^{(-)}  -  \pv ) \cdot
   ( \tilde{\qv}_n  - \qv )
    - ( - 2 \pv_{n}^{(-)} + \tilde{\pv}  + \pv_{n}^{(-)} ) \cdot
     ( \tilde{\qv} -  \qv_n ) 
\end{equation}
\begin{equation}
2 \, \delta S_{L} 
  = 
  ( 2 \pv^{(-)} + \tilde{\pv}^{(-)}  - \pv^{(-)} ) \cdot
   ( \tilde{\qv}  - \qv )
    - (  2 \pv_{n}   + \tilde{\pv}_{n}  - \pv_{n}   ) \cdot  
     ( \tilde{\qv}_n -   \qv_n  )  \;.
\end{equation}
A calculation without difficulties leads to
\begin{eqnarray}  \label{eq-deltaS}
\nonumber
\displaystyle 
& 2\, \delta S =  \delta  S_R + \delta  S_L 
 =   
 \bigl( \tilde{\xv} - \xv \bigr) \wedge 
   \bigl( T_\Gamma \, \tilde{\xv}_{n}  - \xv  
  \bigr) 
    +  \bigl( T_\Gamma \,\tilde{\xv}_{n}  -  T_\Gamma \,\xv_{n}  \bigr) 
      \wedge \bigl( \tilde{\xv}  -  T_\Gamma \, \xv_{n} \bigr)
\\
& 
\displaystyle 
   - \bigl( \tilde{\mbox{\boldmath{$\sigma$}}} 
+ \mbox{\boldmath{$\sigma$}} \bigr) \cdot \bigl( \tilde{\qv} - \qv \bigr) 
  - \tilde{\mbox{\boldmath{$\sigma$}}}_n  \cdot ( \tilde{\qv}_n - \qv \bigr)
    - \mbox{\boldmath{$\sigma$}}_n \cdot ( \tilde{\qv} -  \qv_n  ) 
       + \ooc (\Delta x^3) \;.
\end{eqnarray}
The $\Gamma$-symplectic product 
$(\yv - \xv) \wedge  (\zv -\xv)$ of two infinitesimal displacements
$(\yv - \xv)$ and $(\zv-\xv)$ tangent to $\Sigma$ at $\xv$, with coordinates $(y-x)$ and $(z-x)$,
reduces to the $\Sigma$-symplectic product $(y-x) \wedge (z-x)$ given by (\ref{eq-symplec_prod})
(a choice of $(q,p)$-coordinates in  $\Sigma$ with these properties is always possible, 
see~\cite{Gutzwiller}).
Hence
 letters in  bold font can be replaced by letters in normal font.
The first term in the second line in (\ref{eq-deltaS}) is of third order in $\Delta x$ by the above argument.
One finds
\begin{eqnarray} \label{eq-bold_sympl_prod}
\nonumber
\delta S & = &    
 \frac{1}{2} \bigl( \tilde{x} - x \bigr) \wedge 
   \bigl( T \tilde{x}_{n}  - x  \bigr) 
    +  \frac{1}{2}  \bigl( T \tilde{x}_{n}  -  T  x_{n}  \bigr) 
      \wedge \bigl( \tilde{x}  -  T   x_{n} \bigr) 
\\
&  &
 - \frac{1}{2}\left(  (\tilde{q} - q_n)^2 - ( q - \tilde{q}_n )^2 \right)
    \kappa_n \, \pv_{n}  \cdot \Nv_n   + \ooc (\Delta x^3)
\;,
\end{eqnarray}
where $\kappa_n$ and $\Nv_n$ are the curvature and the normal vector of $\partial \Omega$ at 
the point $\qv_n$ of arc length $q_n$.
Since $x$, $\tilde{x}$, $T x_n$ and $T \tilde{x}_n$ form a parallelogram 
to lowest order (see section~\ref{sec-action_diff}),   
$\tilde{x} - T x_n \simeq  x - T \tilde{x}_n$
and the last term is of higher order in $\Delta x$. Therefore, (\ref{eq-bold_sympl_prod}) reduces
to the canonical invariant expression (\ref{eq-deltaScont2}). 
Note that this result holds for any dimension
of the phase space $\Gamma$.

\end{appendix}

\pagebreak




\begin{thebibliography}{99}

\bibitem{Sieber01} M. Sieber and K. Richter, Phys. Scr. T {\bf 90}, 128 (2001);
M. Sieber, J. Phys. A: Math. Gen. {\bf 35}, L613 (2002)


\bibitem{BGS84} O. Bohigas, M.J. Giannoni, and C. Schmit, Phys. Rev. Lett. {\bf 52}, 
1 (1984)  


\bibitem{Haake} F. Haake, {\it Quantum Signatures of  Chaos}, 2nd edn (Springer, Berlin, 2000) 

\bibitem{Gutzwiller} M.C. Gutzwiller, {\it Chaos in Classical and Quantum Mechanics}
(Springer, New York, 1990)

\bibitem{Berry85} M.V. Berry, Proc. R. Soc. A {\bf 400}, 229 (1985)

\bibitem{Pechukas83} P. Pechukas, Phys. Rev. Lett. {\bf 51}, 943 (1983); F. Haake, 
M. Ku\'{s} and R. Scharf, Z. Phys. B {\bf 65}, 381 (1987)

\bibitem{Agam95} A.V. Andreev, O. Agam, B.D. Simons, and B.L. Altshuler,
Phys. Rev. Lett. {\bf 76}, 3947 (1996);  A.V. Andreev, B.D. Simons,  O. Agam, and B.L. Altshuler,
Nucl. Phys. B, {\bf 482}, 536 (1996)



\bibitem{Argaman93} N. Argaman, F.-M. Dittes, E. Doron, 
J. Keating, A. Kitaev, M. Sieber, and U. Smilansky, Phys. Rev. Lett. {\bf 71}, 
4326 (1993); see also D. Cohen, H. Primack and U. Smilansky, Ann. Phys. {\bf 264}, 108 (1998)

\bibitem{Schanz02} G. Berkolaiko, H. Schanz, and R. Whitney,
Phys. Rev. Lett. {\bf 88}, 104101 (2002)

\bibitem{Marko} M. Turek and  K. Richter, to be published in J. Phys. A; arXiv:nlin.CD/0303053 

\bibitem{Mueller03} S. M\"uller, submitted to Eur. Phys. J. B; arXiv:nlin.CD/0303050 


\bibitem{Gaspard} P. Gaspard, {\it Chaos, Scattering and Statistical Mechanics}
(Cambridge University Press, Cambridge, 1998)



\bibitem{Braun02} P. Braun, F. Haake, and S. Heusler, J. Phys. A: Math. Gen. {\bf{35}},
1381 (2002)

\bibitem{Chernov96} 
N.I. Chernov and C. Haskell, Ergodic Theory and Dyn. Syst. {\bf 16}, 19 (1996);
M. Wojtkowski, Commun. Math. Phys. {\bf 105}, 391 (1986) 

\bibitem{Bowen72} R. Bowen, Amer. J. Math. {\bf 94}, 1 (1972);
W. Parry and M. Pollicott, {\it Zeta Functions and the Periodic Orbit 
Structure of Hyperbolic Dynamics}, Ast\'erisque vol 187-188 (Soci\'et\'e math\'ematique
de France, 1990)

\bibitem{Knieper98} G. Knieper, Ann. Math. {\bf 148}, 291 (1998)

\bibitem{Katok} A. Katok, J.-M. Strelcyn, {\it Invariant Manifolds, Entropy and Billiards:
Smooth Maps with Singularities}, Lecture Notes in Mathematics 1222 (Springer, Berlin, 1986)

 \bibitem{Prange97} R.E. Prange,  Phys. Rev. Lett. {\bf 78}, 
2280 (1997); F. Haake, H.-J. Sommers and J. Weber, J. Phys. A: Math. Gen.
{\bf 32}, 6903 (1999)


\bibitem{Heusler02} P.A. Braun, S. Heusler, S. M\"uller, and F. Haake, 
Eur. Phys. J. B {\bf 30}, 189 (2002)


\bibitem{Mueller} S. M\"uller, Diploma Thesis, Essen (2001)


\bibitem{HODA84} J.H. Hannay
and A.M. Ozorio de Almeida, J. Phys. A: Math. Gen. {\bf 17}, 3429 (1984)



\bibitem{Berkolaiko02}  G. Berkolaiko, H. Schanz, and R. Whitney, arXiv:nlin.CD/0205014

\bibitem{Smilansky02} U. Smilansky and B. Verdene,  J. Phys. A: Math. Gen. {\bf 36}, 3525 (2003)

\bibitem{Richter02} K. Richter and M. Sieber, Phys. Rev. Lett. {\bf 89}, 206801-1
(2002)



\end{thebibliography}
\end{document}